\documentclass[traditabstract]{aa} 
\usepackage{amsmath}
\usepackage{graphicx}
\usepackage{txfonts}

\usepackage{natbib}
\bibpunct{(}{)}{;}{a}{}{,}
\usepackage{float}
\usepackage{calc}
\usepackage{textcomp}
\usepackage{placeins}
\usepackage{color}
\usepackage{rotating} 
\usepackage{lscape}
\usepackage{url}
\usepackage[colorlinks=true,linkcolor=blue,citecolor=blue]{hyperref}
\usepackage{subfig}
\usepackage[all]{draftcopy}
\usepackage{longtable}
\usepackage{array}
\usepackage{threeparttable}
\usepackage{lscape}
\DeclareTextSymbol{\degre}{OT1}{23}
\newcounter{savedfootnote}

\def \microns{{\,$\mu$m}}

\begin{document}
   \title{Identification of galaxies that experienced a recent major drop of star formation}

\author{L.~Ciesla\inst{1,2,3}, D.~Elbaz\inst{1,2}, C.~Schreiber\inst{4}, E.~Daddi\inst{1,2}, and T.~Wang\inst{1,5}.
}

\institute{	
 IRFU, CEA, Universit\'e Paris-Saclay, F-91191 Gif-sur-Yvette, France
 \and
 Universit\'e Paris Diderot, AIM, Sorbonne Paris Cit\'e, CEA, CNRS, F-91191 Gif-sur-Yvette, France
\and
Aix Marseille Univ, CNRS, CNES, LAM, Marseille, France
\and
Leiden Observatory, Leiden University, 2300 RA, Leiden, The Netherlands
\and
Institute of Astronomy, The University of Tokyo, 2-21-1 Osawa, Mitaka, Tokyo, 181-0015 Japan
}	
 
   \date{Received; accepted}

  \abstract
{
Variations of star formation activity may happen on a large range of timescales and some of them are expected to be short, that is, a few hundred Myr.
The study of the physical processes linked to these rapid variations requires for large statistical samples to be able to pinpoint galaxies undergoing such transformations.
Building upon a previous study, we define a method to blindly identify galaxies that underwent, and may still be undergoing, a fast downfall of their star-formation activity, that is, a more than 80\% drop in star-formation rate (SFR) occurring in less than 500\,Myr.
Modeling galaxies' SED with a delayed-$\tau$ star formation history (SFH), with and without allowing an instantaneous SFR drop within the last hundreds Myr, we isolate 102 candidates out of a subsample of 6,680 galaxies classified as star-forming from the UVJ criterion in the ZFOURGE catalogues.
These galaxies are mostly located in the lower part of the SFR-M$_*$ main sequence (MS) and extend up to a factor 100 below it. 
They also lie close to the limit between the passive and active regions on the UVJ diagram, indicating that they are in a transition phase.
We show that the selected candidates have different physical properties compared to galaxies with similar UVJ colors, namely, lower star-formation rates and different stellar masses.
The morphology of the candidates show no preference for a particular type.
Among the 102 candidates, only 4 show signs of an AGN activity (from X-ray luminosity or UV-IR SED fitting decomposition).
This low fraction of AGNs among the candidates implies that AGN activity may not be the main driver of the recent downfall, although timescale differences and duty cycle must be taken into account.
We finally attempt to recover the past position of these galaxies on the SFR-M$_*$ plane, before the downfall of their star-formation and show that some of them were in the starburst region before and are back on the MS.
These candidates constitute a promising sample that need more investigation in order to understand the different mechanisms at the origin of the star formation decrease of the Universe since $z$$\sim$2.
}

  
   \authorrunning{Ciesla et al.}
   \titlerunning{Fast SF decrease from broad-band SED}

   \maketitle

\section{\label{intro}Introduction}
Ten years ago, the discovery of the tight relation linking the star formation rate (SFR) and stellar mass of star-forming galaxies opened a new window in our understanding of galaxy evolution \citep{Elbaz07,Noeske07}.
The main consequence of this relation is that galaxies are forming the bulk of their stars through steady state processes rather than violent episodes of star formation.
This galaxy main sequence (MS) is found to hold up to $z$=4 \citep{Schreiber17} and its normalisation and shape to vary with redshift \citep{Daddi07,Pannella09,Elbaz11,Rodighiero11,Speagle14,Whitaker14,Schreiber15,Gavazzi15,Tomczak16}.
However, what is striking is that the scatter of the main sequence is found to be relatively constant at all masses and over cosmic time \citep{Guo13,Ilbert15,Schreiber15}.
A possible origin of this scatter is that it could be artificially created by the accumulation of errors in the extraction of photometric measurements and/or in the determination of the SFR and/or stellar mass in relation with model uncertainties. 
However, several studies have found a coherent variation of physical galaxy properties such as the gas fraction \citep{Magdis12}, Sersic index and effective radius \citep{Wuyts11}, U-V color \citep[e.g.,][]{Salmi12} suggesting that the bulk of the scatter is related to physics and not measurement and model uncertainties. 
Oscillations of the SFR resulting from a varying infall rate and compaction of star-formation have also been advocated to explain the MS scatter \citep{Tacchella16}.

The path of galaxies on the SFR-M$_*$ plane is still debated.
Do galaxies evolve along the MS, growing in mass?
Do they reach at some point the starburst region and then go back on the MS or quench?
Do they undergo small variations going above and/or below the MS?

Some simulations predict that galaxies undergo some fluctuations of star formation activity resulting in variations of their SFR such as compaction or variations of accretion \citep[e.g.,][]{DekelBurkert14,Sargent14,Scoville16}. 
These variations must be small enough to keep the SFR of the galaxy within the MS scatter.
From an observational point of view, \cite{Elbaz17} showed that some massive compact galaxies exhibiting starburst galaxy properties (short depletion time and high infrared surface density) can be found within the MS.
However they have different morphologies and gas fraction compared to "true" starbursts (above the MS), indicating a different origin, possibly  being late-stage mergers of gas-rich galaxies.
This could be a clue of a possible recent movement of these galaxies from the starburst galaxies region back to the MS. 
However, recent studies pointed toward alternative interpretations to the in situ scenario implied by the MS based on the young ages found for starburst galaxies \citep{daCunha15,Ma15,Mancuso16}.
Thus, there is no consensus on the evolution of galaxies relative to the MS.

To tackle this problem, one has to study the recent star formation history of galaxies.
This information is embedded in their spectral energy distribution (SED).
However, recovering it through SED modeling is complex and subject to many uncertainties and degeneracies.
Although an average SFH of galaxies can be derived assuming that they follow the MS \citep{Heinis14,Ciesla17}, galaxies are expected to exhibit complex SFHs, with short term fluctuations, needing sophisticated SFH parametrizations to model them \citep[e.g.,][]{Lee10,Pacifici13,Behroozi13,Pacifici16}.
However, these models are complex to implement and a large library is needed to be able to model all galaxies properties.
Instead, numerous studies used simple analytical forms to model galaxies SFH \citep[e.g.,][]{Papovich01,Maraston10, Pforr12,Gladders13,Simha14,Buat14,Boquien14,Ciesla15,Abramson16,Ciesla16,Ciesla17}.
Furthermore, SFH parameters are known to be difficult to constrain from broadband SED modeling with a general agreement on the difficulty to constrain the age of the galaxy, here defined as the age of the oldest star, from broad-band SED fitting \citep[e.g.][]{Maraston10,Pforr12,Buat14,Ciesla15,Ciesla17}.
To understand the origin of the scatter of the MS, we need to use an analytical SFH that must able to recover recent variations of the SFR with a precision better than the scatter of the MS itself, i.e. 0.3 dex.
Recently, \cite{Ciesla17} showed that a delayed SFH to which we add a flexibility on the recent SFH provides SFR that are more accurate than those estimated by other typical analytical SFHs ($\tau$ models, delayed, etc...).
This SFH was tested in \cite{Ciesla16}.
Studying a sample of local galaxies from the Virgo cluster, that is galaxies known to have undergone a fast drop of star formation activity due to ram pressure stripping, we showed that the amplitude of the flexibility can be constrained by broadband SED modeling as long as UV rest frame and near-IR data are available.

Relying on these results, we propose a new approach to probe the recent star formation history of galaxies and identify those that underwent a recent ($\leq$500\,Myr) and major ($\geq$80\%) drop of their star-formation activity.
The goal of our study is to pinpoint recent movements in the SFR-M$_*$ plane.
In Sect.~\ref{sample}, we present the selected sample of galaxies drawn from the ZFOURGE survey.
In Sect.~\ref{cig}, we detail the SED modeling code and procedure used in this work.
The definition of the criterion used to select galaxies that underwent a recent and major downfall of their SFR is presented in Sect.~\ref{id}.
The properties of the selected candidates are presented and studied in Sect.~\ref{properties}, Sect.~\ref{morphology}, and Sect.~\ref{agn}.
The possibility of probing movements inside the star-forming main sequence is discussed in Sect.~\ref{variations}.

Throughout the paper, we use the WMAP7 cosmological parameters \citep[$H_0=70.4$\,km\,Mpc$^{-1}$\,s$^{-1}$, $\Omega_0=0.272$,][]{Komatsu11}.
SED fitting is performed assuming an IMF of \cite{Salpeter55}, but the results of this work are found to be robust against IMF choice.
\section{\label{sample}The sample}	

The aim of this work is to identify galaxies experiencing a recent and fast decrease of their star-formation activity.
In \cite{Ciesla16}, we showed that the method we use in this study is sensitive to rapid downfall of star-formation that occurred in less than $\sim$500\,Myr.
Since the downfall considered here happens in a very short time, we need a large statistics to be able to catch the responsible physical processes in the act.
Another key to perform this study is to have a good and large sampling of the UV-NIR SED of galaxies.
For these two reasons, we draw our sample from the ZFOURGE/GOODS-South catalogue \citep{Straatman16} which contain intermediate bands observations exactly in the key spectral domain where we expect the SED to be sensitive to one star formation history (SFH) or the other.

From the ZFOURGE/GOODS-South parent sample, we select galaxies with a redshift comprised between 0.5 and 3.5 and a stellar mass higher than 10$^9$\,M$_{\odot}$ \citep[as published in][]{Kriek09}.
This corresponds to a final sample of 7,471 galaxies for which we will model the UV to IR SED, and among which 6,680 galaxies are classified as star-forming following a UVJ color selection.
The photometric bands used for the fits are listed in Table~\ref{bands}.
The photometric catalogues were extensively described in \cite{Straatman16}, and references therein, we refer the reader to these papers for further information on the photometry and catalogue building.
We impose of a minimum of three for the signal-to-noise ratio of the flux of each photometric band.
In the rest of the paper, we will use the U-V and V-J colors provided by \cite{Straatman16} as we used them to select our sample.
However, the stellar masses (and SFR) used in this work result from the SED fitting presented in Sect.~\ref{fitproc}.

\begin{table*}
	\centering
	\caption{Photometric coverage of the 7,471 ZFOURGE galaxies used in this work. Only detections are considered in each band. }
	\begin{tabular}{l l l }
	\hline\hline
	Instrument & $\lambda$ (\microns) & Reference \\ 
       \hline	
	U						&0.36 											& \cite{Nonino09}\\
	U38, B, V, Rc			&0.38, 0.441, 0.551, 0.659						& \cite{Hildebrandt06}, \cite{Erben05}\\
       \hline
	Intermediate				&0.427, 0.445, 0.505, 0.527,  						& \\	
	 bands					&0.550, 0.574, 0.598, 0.624,  						& \cite{Cardamone10}\\	
							&0.651, 0.679, 0.738, 0.767, 						& \\	
							&0.797, 0.856									&\\	
       \hline
	HST/ACS 				&0.435, 0.606, 0.775, 0.814, 0.85					&\cite{Giavalisco04}, \cite{Koekemoer11}\\
	HST/WFC3				&1.05, 1.25, 1.40, 1.60							&\cite{Brammer12}, \cite{Grogin11}, \cite{Koekemoer11}\\	
       \hline
	J, H, Ks					&1.24, 1.65, 2.17 								&\cite{Retzlaff10}, \cite{Wuyts08}, \cite{Hsieh12}\\	
       \hline
	\textit{Spitzer}/IRAC		&3.6, 4.5, 5.8, 8									&\cite{Ashby13}, \cite{Dickinson03}\\			
	 \textit{Spitzer}/MIPS		&24												&\\
       \hline
	\textit{Herschel}/PACS	&100, 160										&\cite{Elbaz11}, \cite{Lutz11} \\
	\textit{Herschel}/SPIRE	&250, 350, 500									&\cite{Oliver12} \\
	\hline
	\label{bands}
	\end{tabular}
\end{table*}
\section{\label{cig}Broadband SED modeling}

	\subsection{The CIGALE code}

	CIGALE\footnote{The code is publicly available at: \url{http://cigale.lam.fr/}.}  (Code Investigating GALaxy Emission) is  a SED modeling software package  that  has  two functions:  a SED modeling  function and a SED fitting function \citep[][Boquien et al., in prep]{Roehlly14}.   
	Even though the philosophy of CIGALE, originally presented in \cite{Noll09}, remains, the code has been rewritten in Python and additions have been made in order to optimize its performance and broaden its scientific applications. 
	The  SED modeling  function of CIGALE allows the building of galaxy SEDs from UV to radio by assuming a combination of modules which model the star formation history (SFH) of the galaxy, the stellar emission from stellar population models \citep{BruzualCharlot03,Maraston05}, the nebular lines, the attenuation by dust \citep[e.g.,][]{Calzetti00}, the IR emission from dust \citep{DraineLi07,Casey12,Dale14}, the AGN contribution \citep{Fritz06,Ciesla15}, and the radio emission.  
	CIGALE builds galaxy SEDs taking into account the balance between the energy absorbed by dust in UV-optical and reemitted in the IR.

	With the SED fitting function of CIGALE, these modeled SEDs are then integrated into a set of filters to be compared directly to the observations.
	For each parameter, a probability distribution function (PDF) analysis is made. 
	The output value is the likelihood-weighted mean value of the PDF and the error associated is the likelihood-weighted standard deviation. 
	We use CIGALE to derive the physical properties of galaxies such as stellar masses, instantaneous SFRs, dust attenuation, SFH characteristics, taking into account panchromatic information from the SED.
	To perform the SED fitting on the ZFOURGE/GOODS-South sample, we rely on photometric redshifts.
	 For the sample used in this work, that is considering the redshift and stellar mass selections, the precision of these redshifts is $\sim$3$\%$.
	
	In CIGALE, the assumed SFH can be  handled in two different ways. 
	The first is to model it using simple analytic functions such as exponential forms or delayed-$\tau$ SFH \citep[see for instance,][]{Ciesla16,Ciesla17}.
	The second is to provide more complex SFHs, such as  those provided by  semi-analytical models (SAM) and simulations \citep{Boquien14,Ciesla15}, an approach which we will not use in the present study.   

	\subsection{\label{fitproc}Modeling and fitting the SEDs}	
	
		\subsubsection{Star formation history modeling}
	
\begin{figure}
  			\includegraphics[width=\columnwidth]{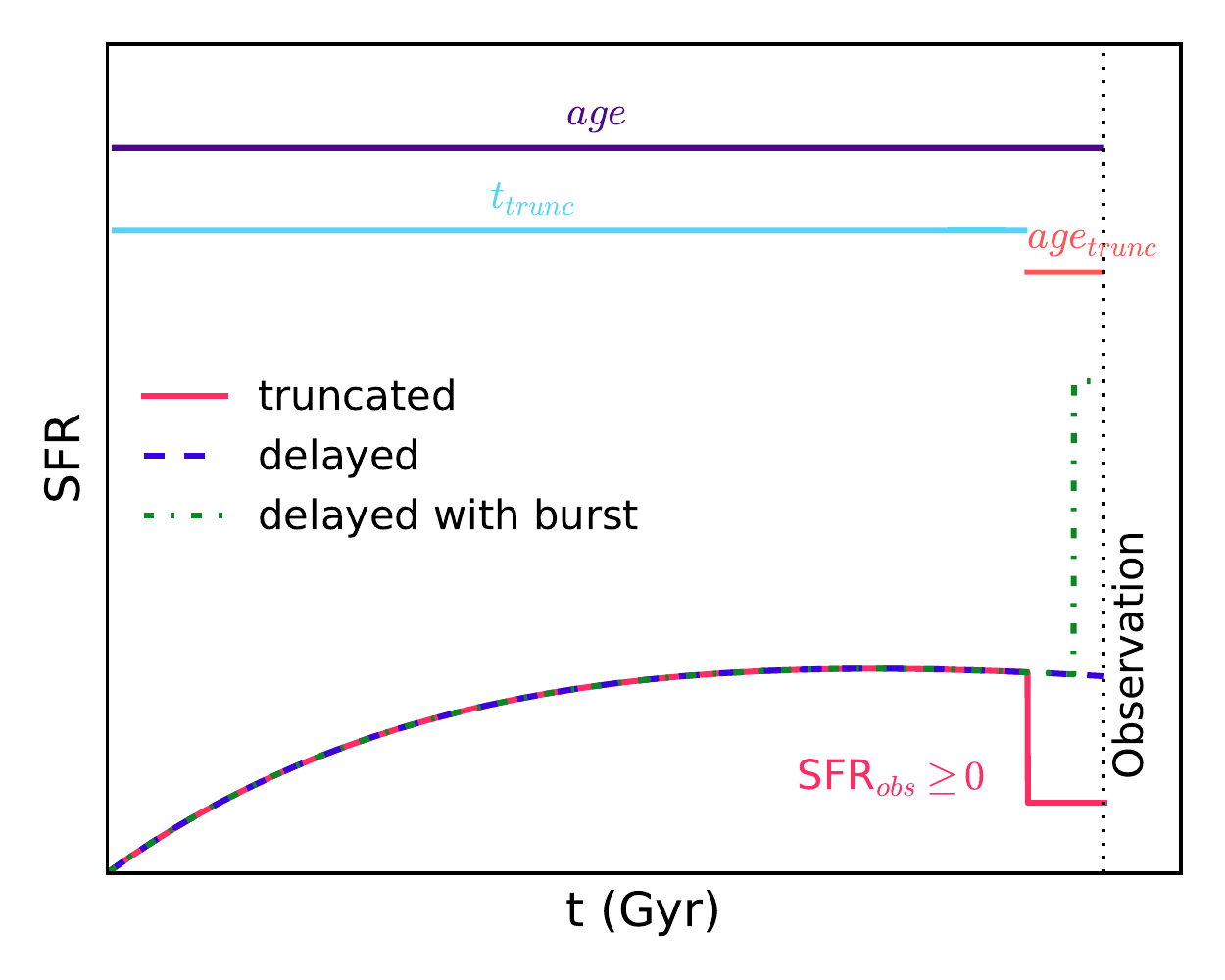}
  			\caption{ \label{sfh_schema} Illustration of the SFH implemented in CIGALE. The purple dashed line represents a normal delayed-$\tau$ SFH with $\tau_{main}$=10\,Gyr, without truncation or burst. The red solid line is the SFH showing a truncation modeling a rapid star-formation downfall and the green dashed-dotted line is a delayed-$\tau$ SFH on top of which we add a star formation burst.}
		\end{figure}

		To identify galaxies that underwent recent important variations in their SFH from broadband SED fitting, we use the SFH implemented in CIGALE by \cite{Ciesla16} to model such a rapid decrease of the SFR:
		\begin{equation}	
    		\mathrm{SFR}(t) \propto
    		\begin{cases}
    	  		t \times exp(-t/\tau_{main}), & \text{when}\ t\leq t_{trunc} \\
    	  		r_{\rm{SFR}} \times \mathrm{SFR}(t=t_{trunc}), & \text{when}\ t>t_{trunc} \\
    		\end{cases}
		\end{equation}	
	
		\noindent where $t_{trunc}$ is the time at which the star formation is instantaneously affected, and $r_{\rm{SFR}}$ is the ratio between SFR$(t>t_{trunc})$ and SFR$(t=t_{trunc})$:
	
		\begin{equation}
			r_{\rm{SFR}} = \frac{\mathrm{SFR}(t>t_{trunc})}{\mathrm{SFR}(t_{trunc})}.
		\end{equation}
		\noindent 	Fig.~\ref{sfh_schema} shows an example of the truncated SFH \citep[see Fig.~3 of ][to see the impact of each parameter on the shape of the SED]{Ciesla16}.
		
		This SFH has been presented and tested intensively in \cite{Ciesla16} where we showed that the $age_{trunc}$ parameter ($age_{trunc} = age - t_{trunc}$) cannot be constrained from broadband SED fitting.
		However, the $r_{\rm{SFR}}$ parameter is well-constrained especially in galaxies that were relatively active before the SFH variation.
		Comprised between 0 et 1, this parameter handles the strength of the rapid star-formation downfall in the SFH.
		Thus, we empirically showed in \cite{Ciesla16} that a star-forming galaxy with an output value of $r_{\rm{SFR}}$ lower than 0.2-0.3 experienced a rapid decrease of its star formation activity in the last $\sim$500\,Myr.
		However, when allowing $r_{\rm{SFR}}$ values larger than 1, it is possible to add a recent burst in the star formation activity of the galaxies modeling starburst galaxies.
		Thus, with this flexible SFH we are able to model a large variety of recent SFH \citep{Ciesla17}.

		\subsubsection{SED fitting procedure}	
		
		To model galaxy SEDs, we used the stellar population models of \cite{BruzualCharlot03} and the attenuation law of \cite{Calzetti00}.
		The energy attenuated in UV-optical is re-emitted in the IR (mostly 8 to 1000\microns) through the models of \cite{Dale14}.
		However, 71\% of the galaxies are not detected by either Spitzer-MIPS or Herschel-PACS/SPIRE. 
		To apply the same procedure with the same constraints to all galaxies, and to prevent the fit quality to be too strongly driven by the IR, we only fit the UV-to-NIR data of the galaxies, i.e. up to \textit{Spitzer}/IRAC4.
		We thus fix the shape of the dust emission since we do no study the dust properties of these galaxies.
		
		To identify galaxies that underwent a rapid decrease of their star formation activity, we perform the fit twice, once using a normal delayed-$\tau$ SFH and the second time using our truncated SFH.
		The input parameters used to perform the SED fitting are presented in Table~\ref{fitparam}.
		Even if we only consider star-forming galaxies, the defined parameter grids enable to fit both active and passive galaxies in case of contamination by passive galaxies.

		\begin{table}
			\centering
			\caption{Galaxy parameters used in the SED fitting procedures.}
			\begin{tabular}{l c }
	 		\hline\hline
			Parameter & Value \\ 
			\multicolumn{2}{c}{Delayed SFH}\\  
			\hline
			$age$ (Gyr) & 0.5, 1, 2, 3, 4, 5, 6, 7, 8, 9, 10	\\
			$\tau_{main}$  (Gyr) & 0.1, 0.3, 0.5, 1, 2, 3, 4, 5, 6, 7, 8, 9, 10, 15, 20\\
			\hline
			\multicolumn{2}{c}{Trunc. delayed-$\tau$ SFH}\\  
			\hline
			$age$ (Gyr) &  0.5, 1, 2, 3, 4, 5, 6, 7, 8, 9, 10	\\
			$\tau_{main}$  (Gyr) & 0.3, 0.5, 1, 2, 3, 4, 5, 6, 7, 8, 9, 10, 15, 20\\
			$age_{trunc}$ (Myr) & 100, 400\\
			$r_{\rm{SFR}}$ 	& 0., 0.01, 0.02, 0.03, 0.04, 0.05,	\\ 
		 			&  0.1, 0.15, 0.2, 0.25, 0.3, 0.35, 0.4,	 \\
		 			&  0.5, 0.6, 0.7, 0.8, 0.9, 1., 3., 6., 12.\\
			\hline
			\multicolumn{2}{c}{Dust attenuation}\\  
			\hline
			$E(B-V)_*$	&  0.01, 0.1, 0.2, 0.3, 0.4, 0.5, 0.6,\\
						&  0.7, 0.8, 0.9, 1.0, 1.1, 1.2, 1.3	\\
			\hline		
			\multicolumn{2}{c}{Dust template: \cite{Dale14}}\\  
			\hline
			$\alpha$	& 2.5  \\
			\hline
			\# of models & 73920\\
			\hline
			\label{fitparam}
			\end{tabular}
		\end{table}

		\subsubsection{Constraints on the output parameters: mock analysis}
		
		To understand what are the output parameters of the SED fitting procedure that we can trust and use in our study, i.e. what are the parameters that are constrained with the photometric coverage available for the ZFOURGE sample, we perform a mock analysis.
		This procedure is thoroughly presented in several works of the literature \citep[e.g.,][]{Giovannoli11}. 
		We briefly describe here the main steps.
		We first apply our SED fitting procedure on the whole real catalogue in order to retrieve the best fit SED for each individual source.
		Each best fit SED is then convolved with the set of filters available for the ZFOURGE sample and a noise is added to the mock flux, randomly chosen in a Gaussian distribution with a $\sigma$ equal to the observed error.
		The original error of each filter is associated to the perturbed flux.
		We thus have a mock photometric catalogue for which we know all the associated physical parameters from the output of the first run.
		A more detailed description on the mock tests implemented in CIGALE can be found in \citep{Giovannoli11} and \citep{Boquien12}.
		
\begin{figure*}
  		\includegraphics[width=\textwidth]{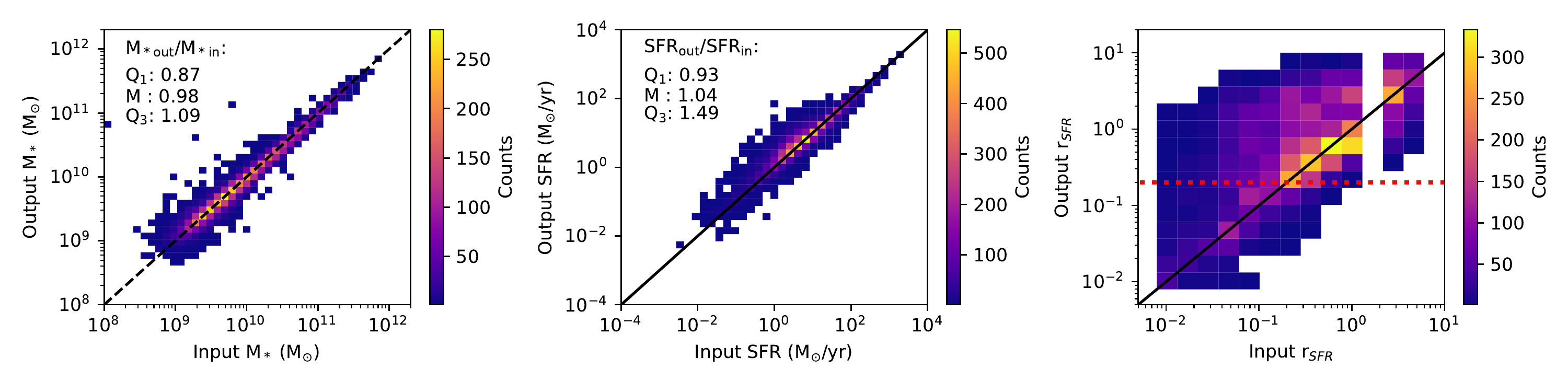}
  		\caption{ \label{mock}  Comparison between the true parameters of the mock SEDs and the results from the SED modeling with CIGALE. A parameter is well-constrained when there is a one-to-one relation between the two. Statistics on the ratio between the output and the input parameters are provided for the stellar mass and SFR: the first quartile of the ratio, the median, and the third quartile. The red dashed line in the $r_{\rm{SFR}}$ panel indicates the threshold adopted in this study.} 
		\end{figure*}

		Then, we apply our SED modeling procedure a second time, but on the mock catalogue and compare the output results to the true values of each parameter we want to use in our study.
		These comparisons are shown in Fig.~\ref{mock} for the stellar mass, SFR, and $r_{\rm{SFR}}$.
		As expected given the good coverage of the rest-frame NIR, the stellar mass is well recovered and thus reliable.
		Although showing a larger dispersion for values between 0.1 and 10\,M$_{\odot}$/yr, the SFR is relatively well constrained with a relation close to the one-to-one line.
		The results of the mock analysis for  $r_{\rm{SFR}}$ show that the parameter is, on average, always overestimated.
		We discuss below that this is not affecting our analysis and the identification of galaxies using this parameter since this overestimation will just make our chosen threshold even more conservative (see Section~\ref{id}).

\section{\label{id}Identifying galaxies with a recent downfall of star formation}

\subsection{Selection criterion}

\begin{figure}
  	\includegraphics[width=\columnwidth]{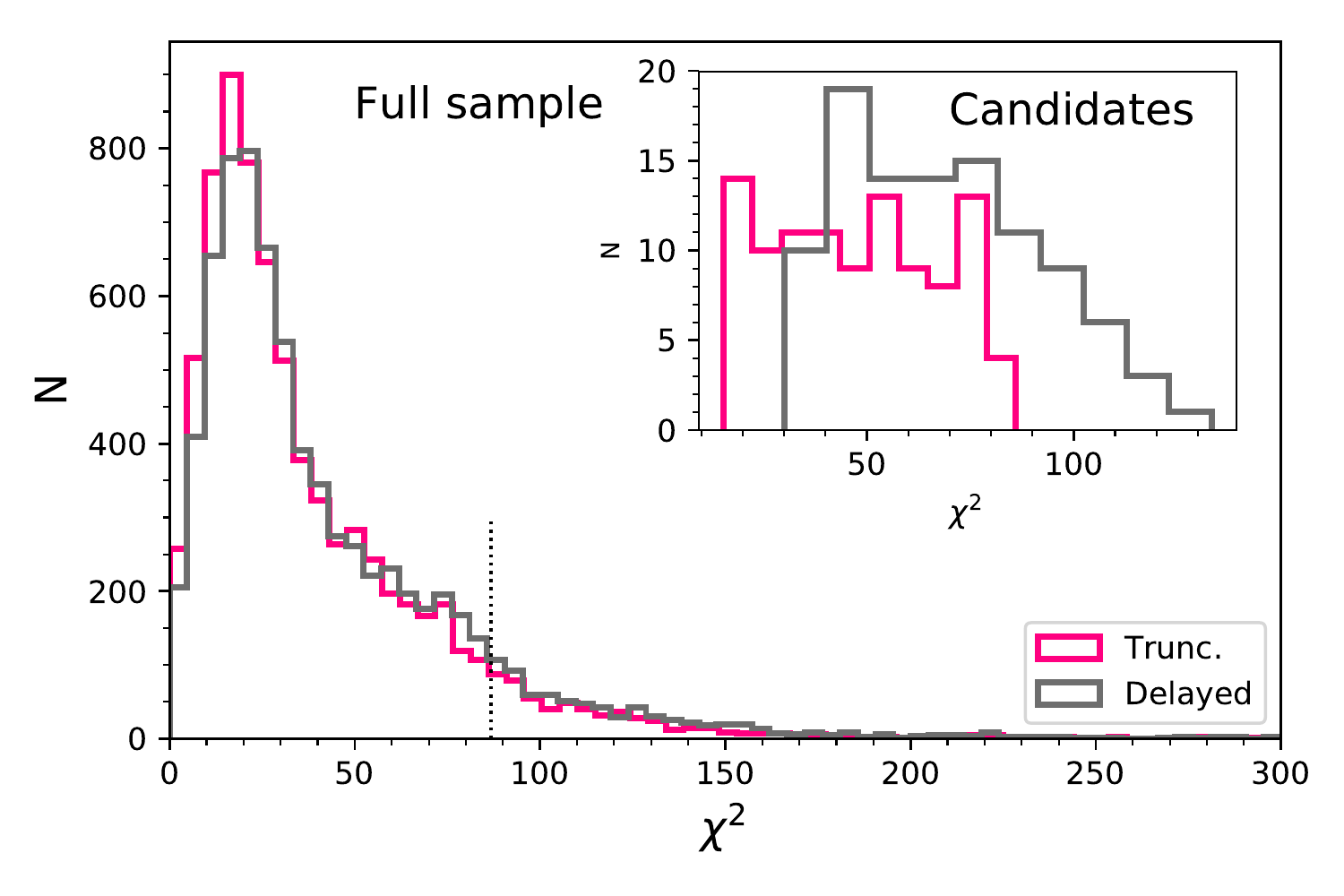}
  	\caption{ \label{chi2} $\chi^2$ distribution obtained with the delayed-$\tau$ and the truncated SFH for the whole sample and for the candidates (inset panel). The dotted line in the main panel indicates the cut made to eliminate the catastrophic fits.}  
\end{figure}

We consider only galaxies defined as star-forming from the UVJ diagram criterion of \cite{Whitaker11} \citep[see also][]{Tomczak14,Spitler14,Straatman14}.
Star formation histories of sources that underwent a rapid decrease of their SFR ($\leq$500\,Myr) are not well fitted by models assuming a normal delayed-$\tau$ SFH, or a exponentially-decreasing one, whereas the truncated SFH proposed manages to reproduce the UV-NIR SED \citep{Ciesla16,Ciesla17}.
We thus expect that the $\chi^2$ obtained with a truncated SFH are significantly lower than the ones from the delayed-$\tau$ SFH for galaxies with a recent decrease of SFR as shown in \cite{Ciesla16}.
In Fig.~\ref{chi2}, we show the distribution of $\chi^2$ obtained using both SFH.
However, the two SFH have different number of degree of freedom.
To take this into account we compute the Bayesian Information criterion (BIC) for each SFH, using the definition:
\begin{equation}
	\mathrm{BIC} = \chi^2 + k \times \ln(n),
\end{equation}
\noindent with $\chi^2$ the non-reduced $\chi^2$ of the fit, $k$ the number of degree of freedom, and $n$ the number of photometric bands fitted.
We then calculate $\Delta$BIC the difference between BIC$_{delayed}$ and BIC$_{trunc}$.
If $\Delta$BIC is larger than 2, then there is a significant difference between the two fits \citep{Liddle04,Stanley15}, with the truncated SFH providing a better fit of the data than the delayed-$\tau$SFH.
However, we impose a $\Delta$BIC$\geq$10 to be conservative.

The second criterion is based on the value of  $r_{\rm{SFR}}$.
Indeed, from the study of Virgo cluster galaxies undergoing ram pressure stripping from the intra-cluster medium, we showed in \cite{Ciesla16} that these galaxies, that underwent a rapid downfall of SFR, have an  $r_{\rm{SFR}}$ of 0.3 and lower.
Here, we consider that galaxies with $r_{\rm{SFR}}\leq$0.2 are possible candidates, to be more conservative.
Although calibrated from a study of galaxies well-known for undergoing a fast drop of their star formation, the value of 0.2 is mostly arbitrary.
However, the goal of this study is not to select a statistically complete sample to derive properties but rather to use  broadband SED fitting to identify rapid phenomena.
This threshold translates in a decrease of the SFR of at least 80$\%$ in the last few hundreds Myr according to  \cite{Ciesla16} where we tested the sensitivity of the truncated SFH in terms of the strength of the star-formation downfall and its timescale.
However, the overestimation of the true $r_{\rm{SFR}}$ value revealed by the mock analysis (Fig.~\ref{mock}) does not affect our criterion but rather goes in the sense of strengthening it since an output value of 0.2 would be an overestimation of the true one.
Finally, we impose a last criterion to reject galaxies with bad fits for both SFH: we reject the 8$\%$ galaxies with the largest $\chi^2_{trunc}$ to remove galaxies with catastrophic fits.
Quantitatively, we exclude galaxies with $\chi^2_{trunc}$$>$88.

\begin{figure}
  	\includegraphics[width=\columnwidth]{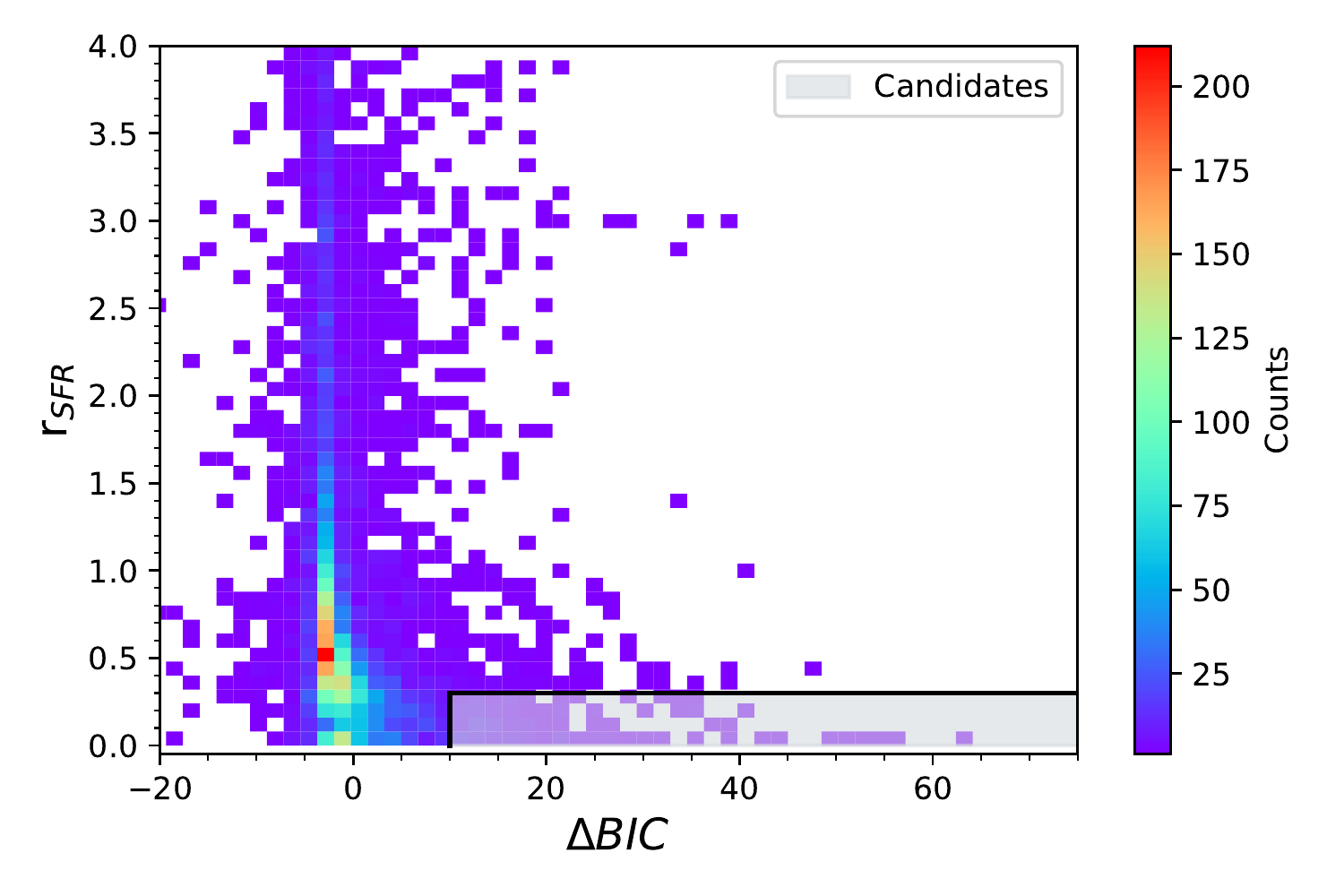}
  	\caption{ \label{rsfr_rchi2} $r_{\rm{SFR}}$ output values as a function of $\Delta$BIC=BIC$_{delayed}$ - BIC$_{trunc}$ for all the star-forming galaxies. The candidates sample is indicated by the gray filled region. For clarity the x-axis is limited to [-20;75] but a few galaxies have higher $\Delta$BIC.}
\end{figure}

From these three criteria, we identify galaxies as candidates if:
\begin{equation}
	\Delta \mathrm{BIC} \geq 10~\&~r_{\rm{SFR}}\leq0.2~\&~\chi^2_{trunc}\leq 88.
\end{equation}
The result of this selection is shown in Fig.~\ref{rsfr_rchi2} where we see that the large majority of the galaxies (98.5\%) lie in the region where $\Delta$BIC$<$10 or $r_{\rm{SFR}}$$>$0.2.
We thus identify 102 sources satisfying our criterion.
The distribution of the $\chi^2$ obtained with the two SFH are shown in the inset panel of  Fig.~\ref{chi2}.

\begin{figure*}
 \center{ 
  	\includegraphics[width=6cm]{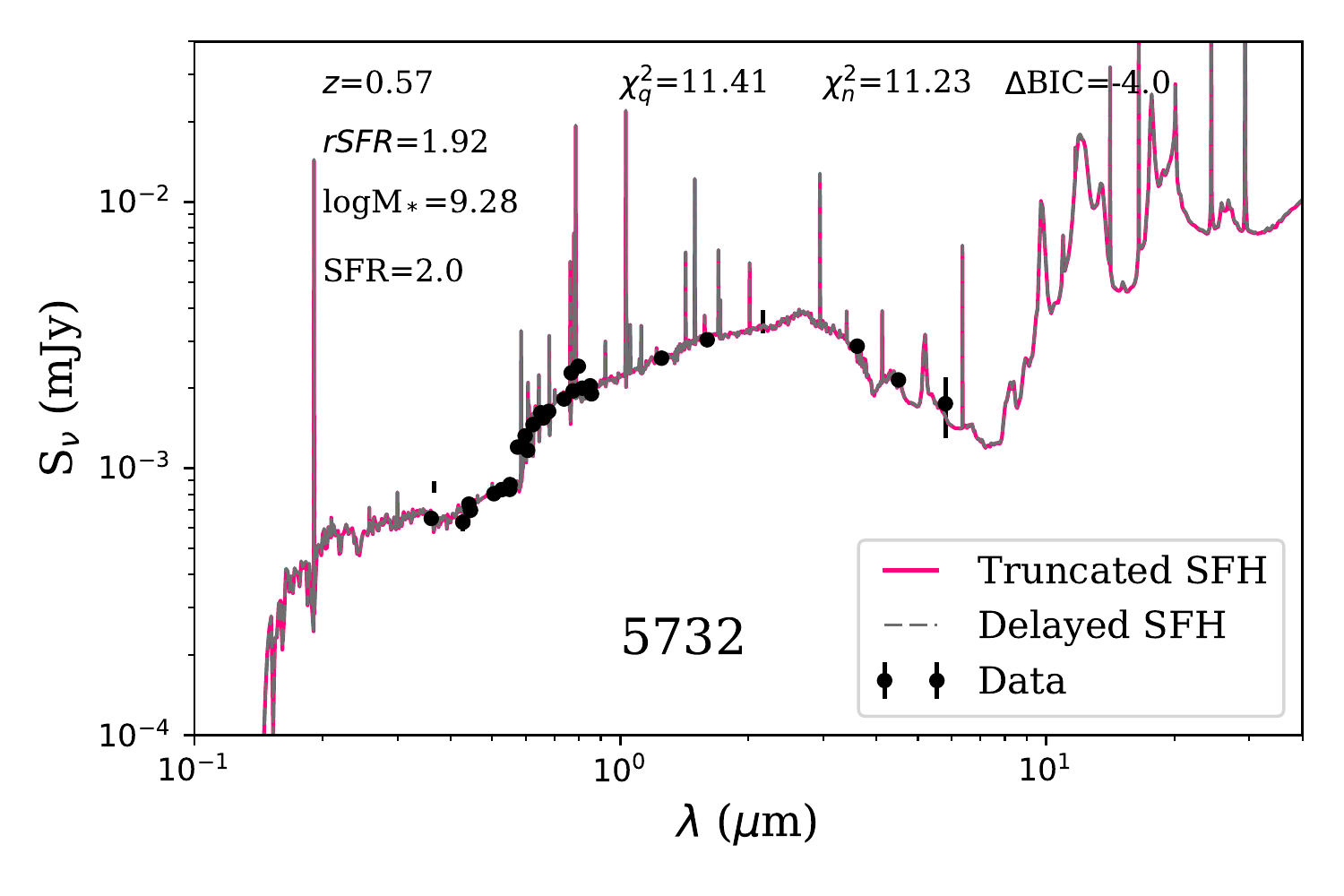}
	\includegraphics[width=6cm]{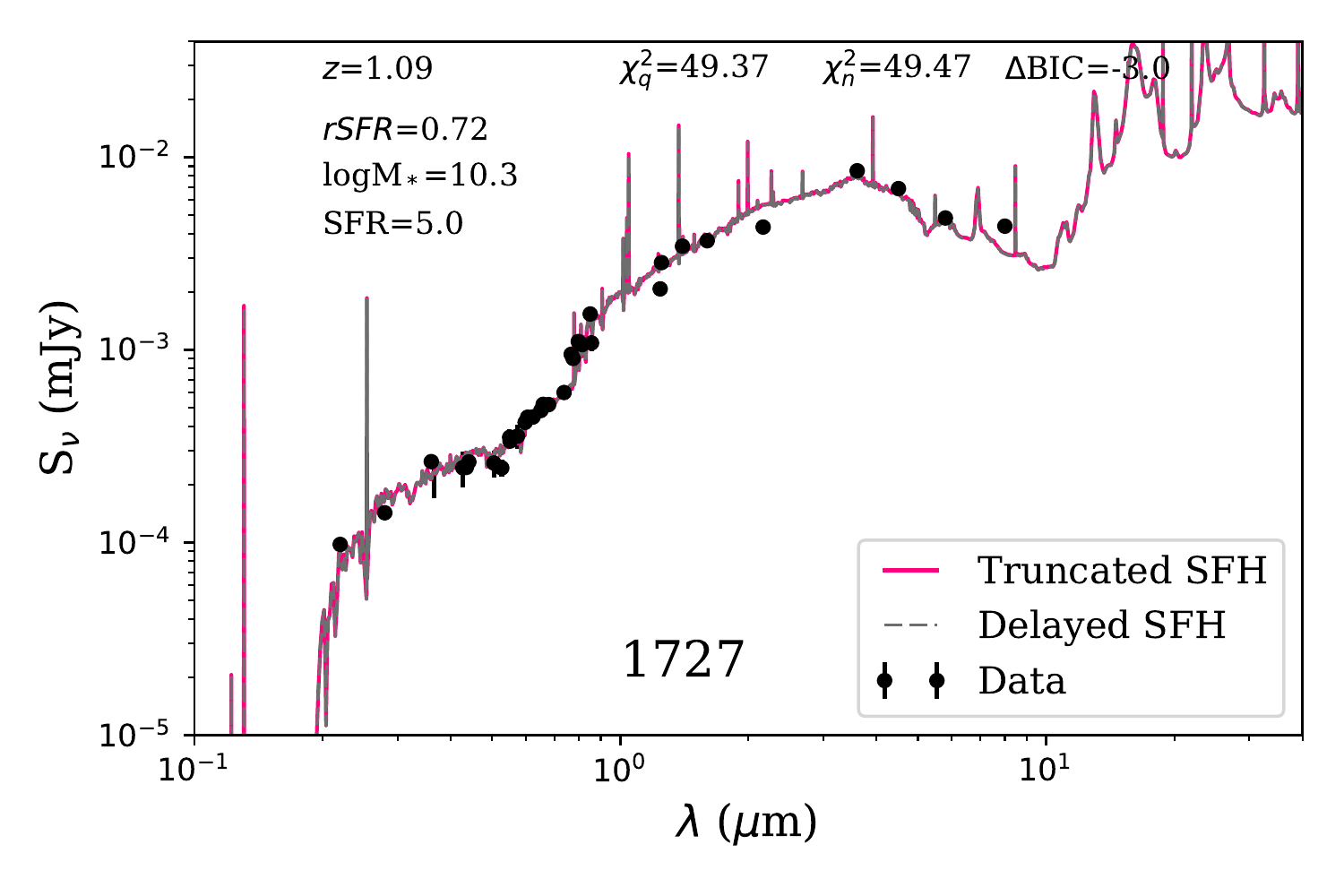}
	\includegraphics[width=6cm]{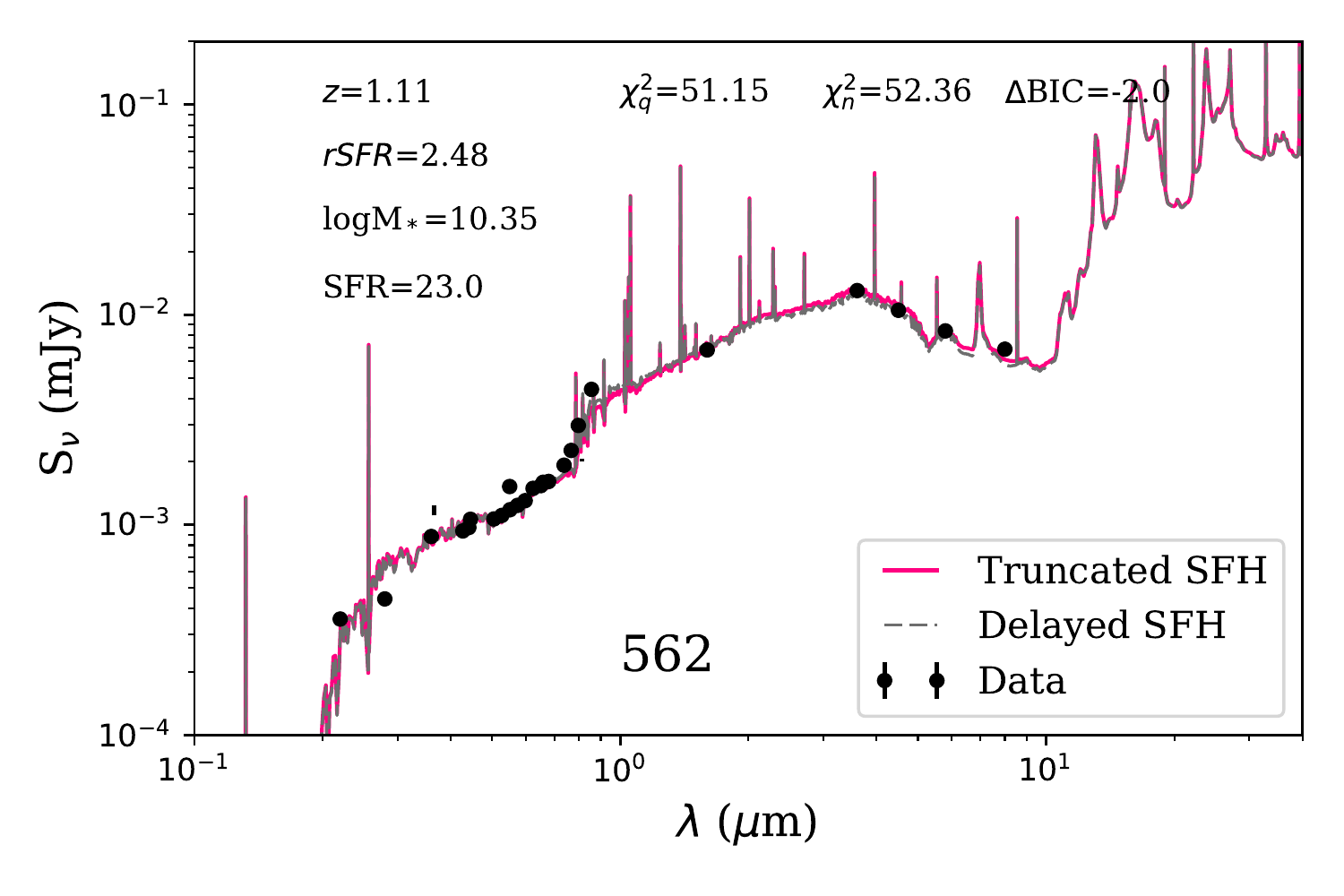}\\
  	\includegraphics[width=6cm]{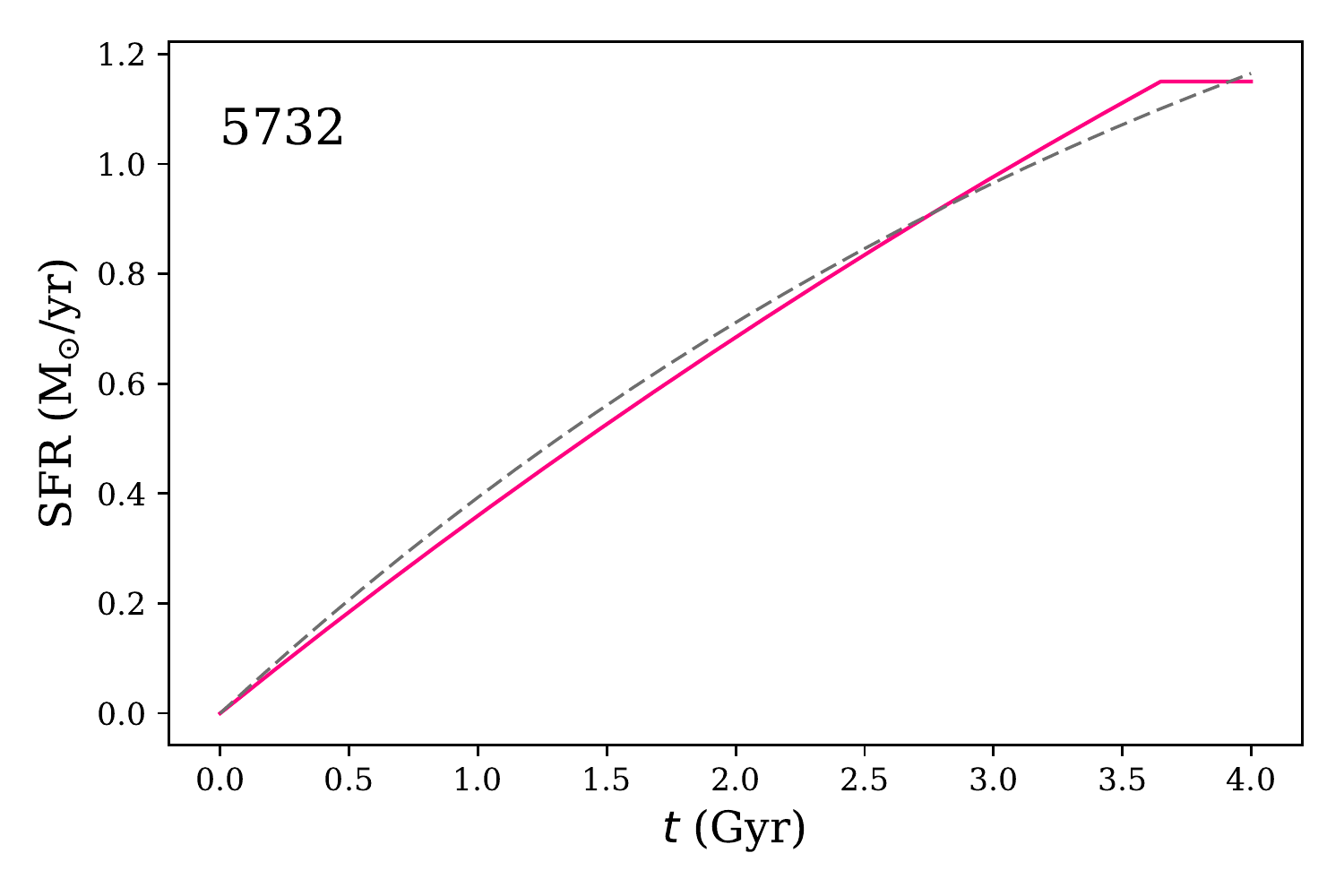}
	\includegraphics[width=6cm]{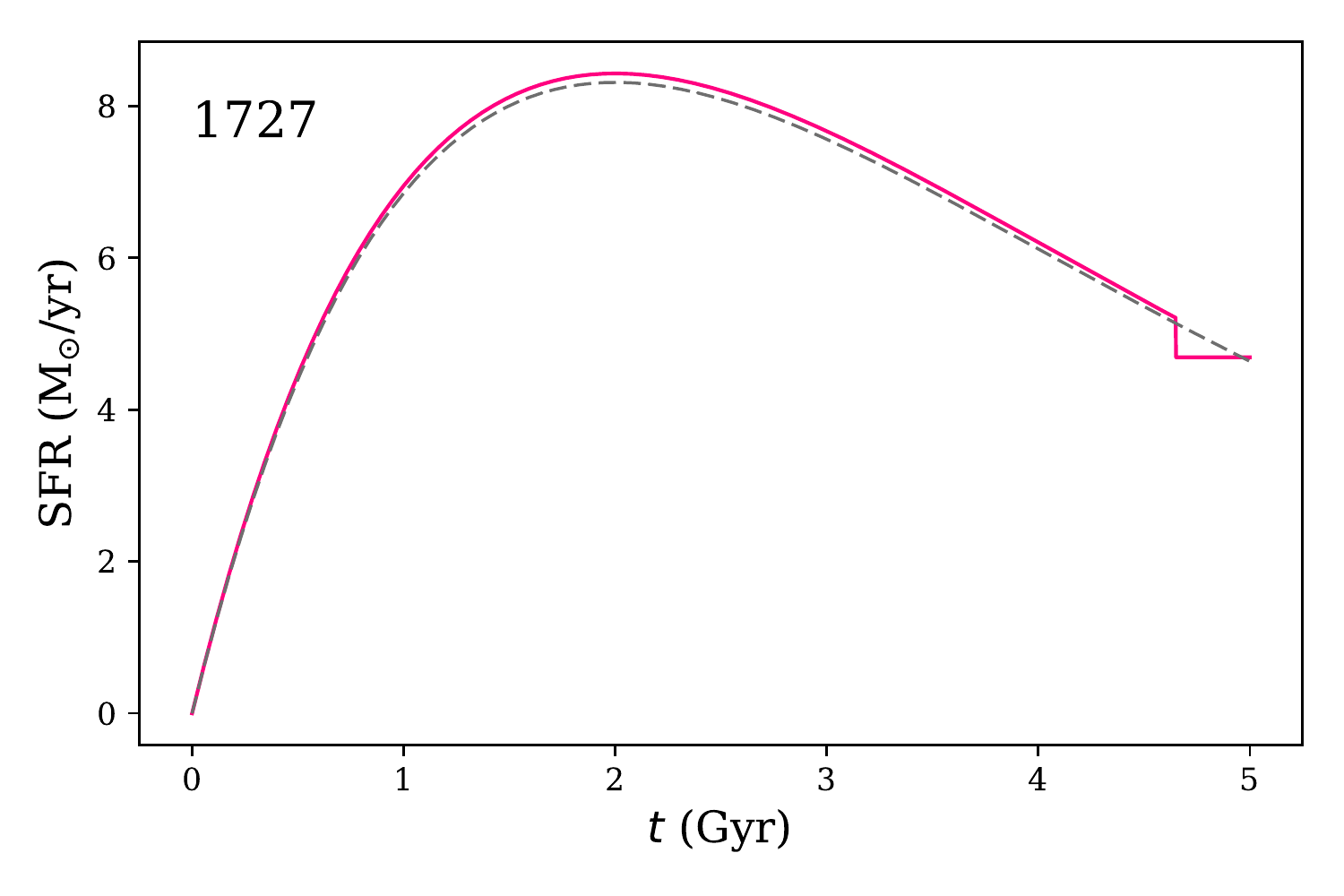}
	\includegraphics[width=6cm]{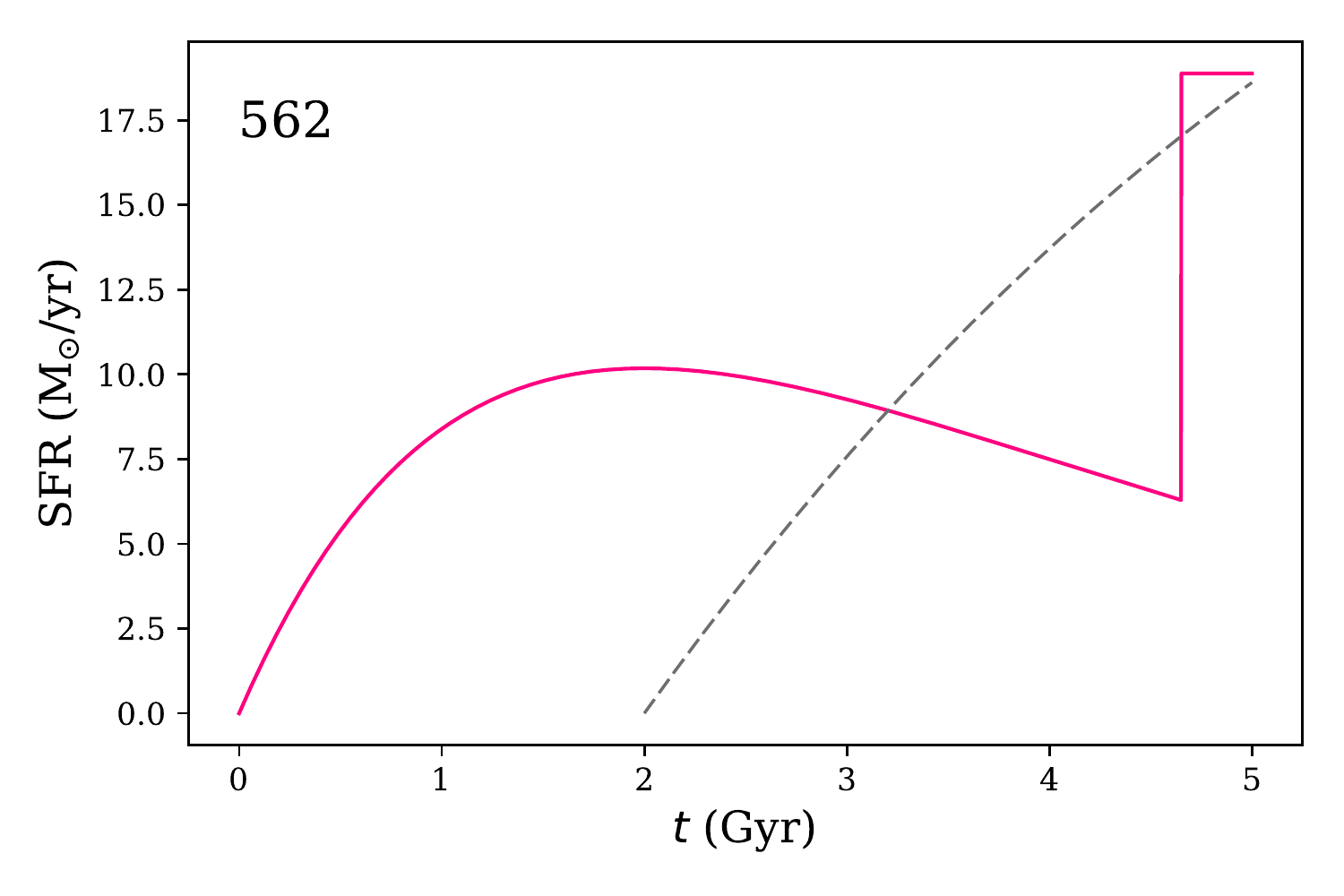}\\	
	\includegraphics[width=6cm]{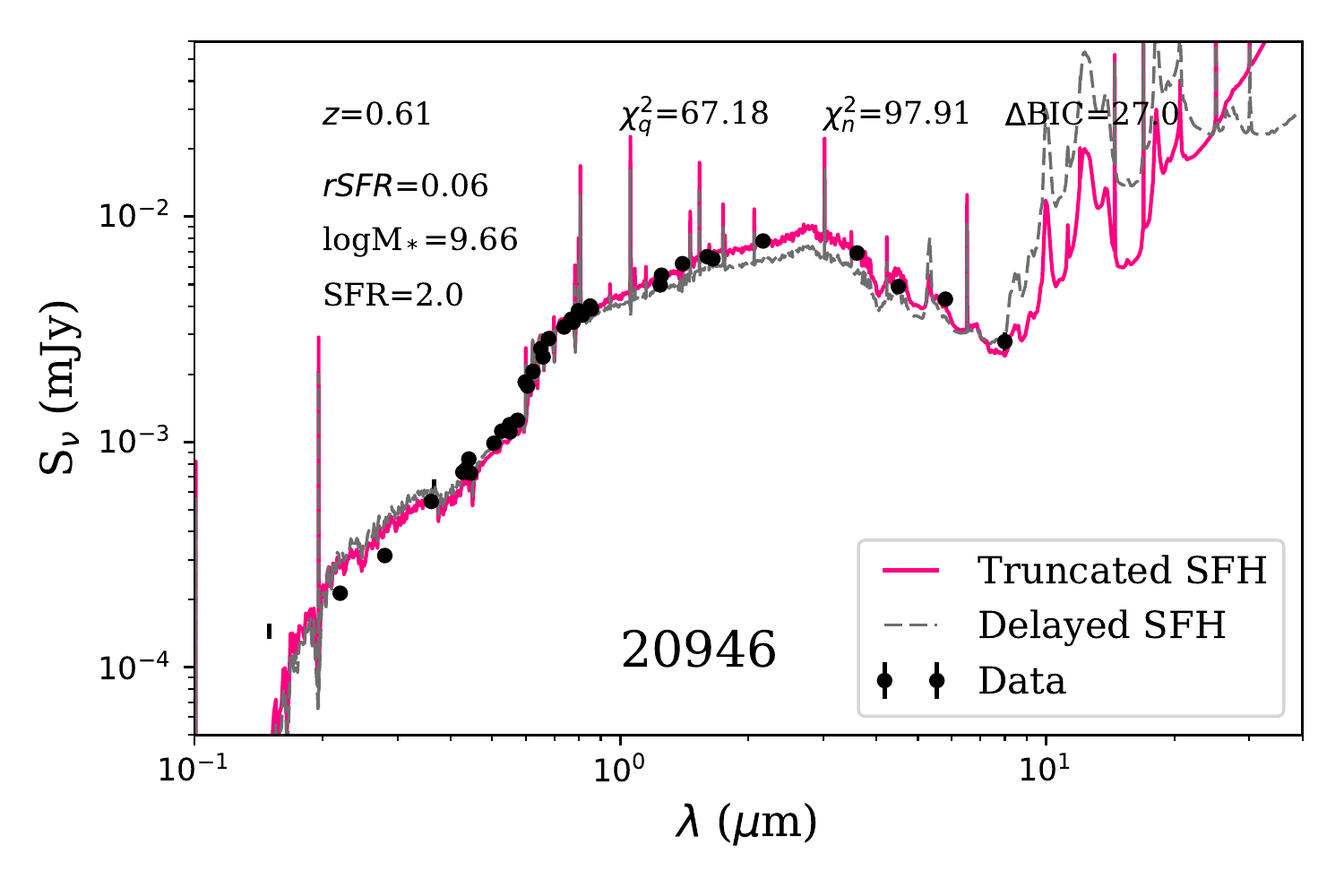}
	\includegraphics[width=6cm]{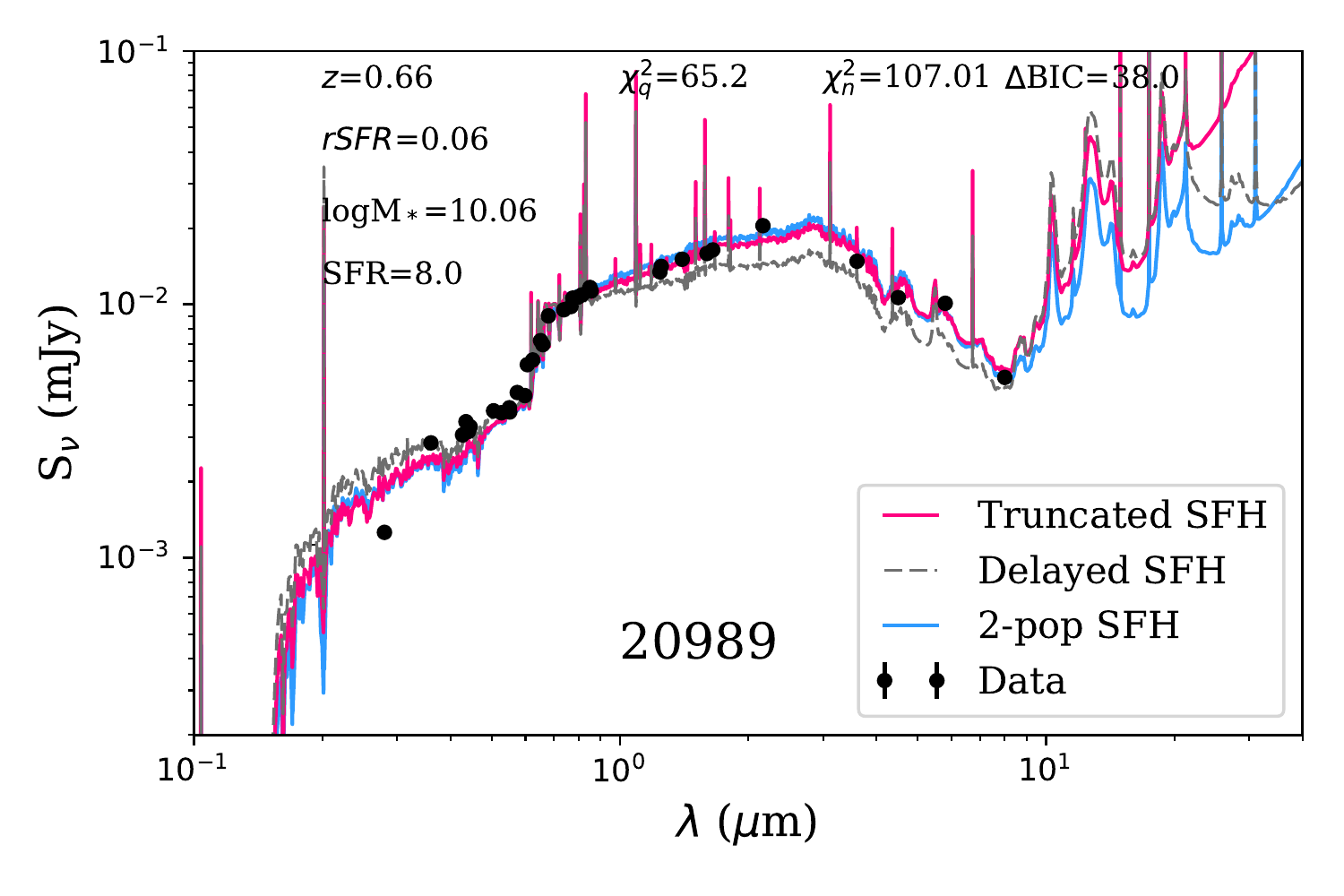}
	\includegraphics[width=6cm]{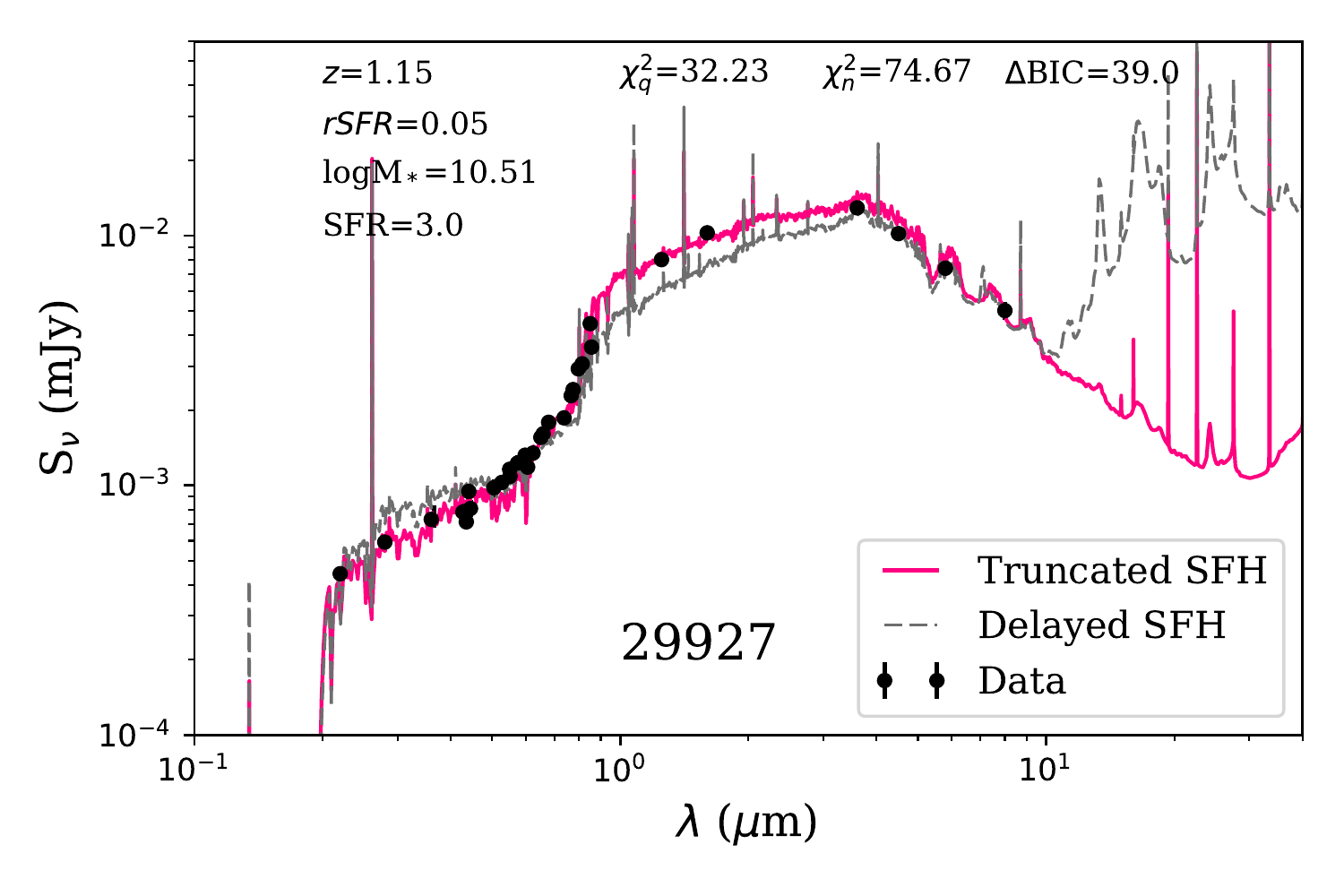}\\
	\includegraphics[width=6cm]{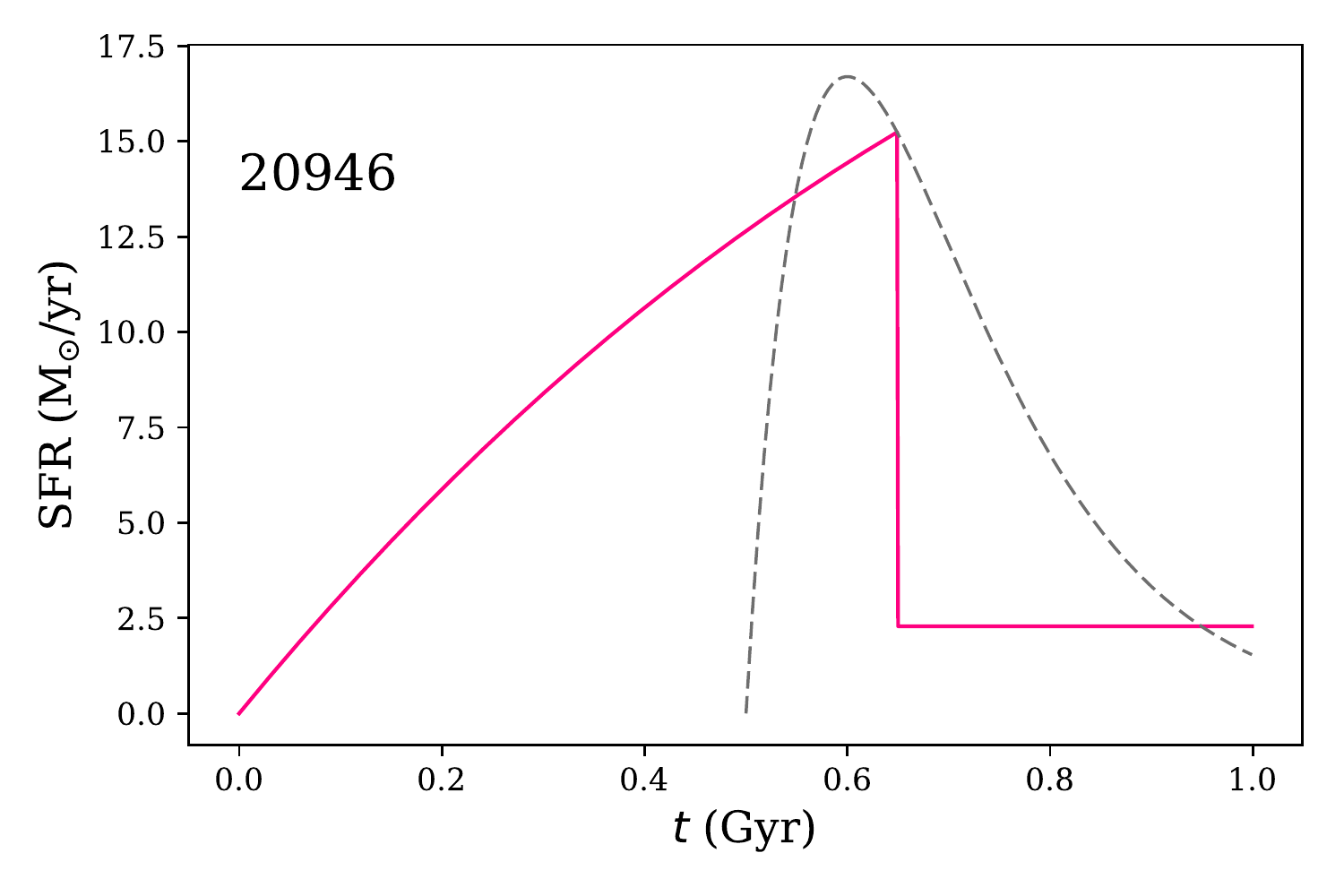}
	\includegraphics[width=6cm]{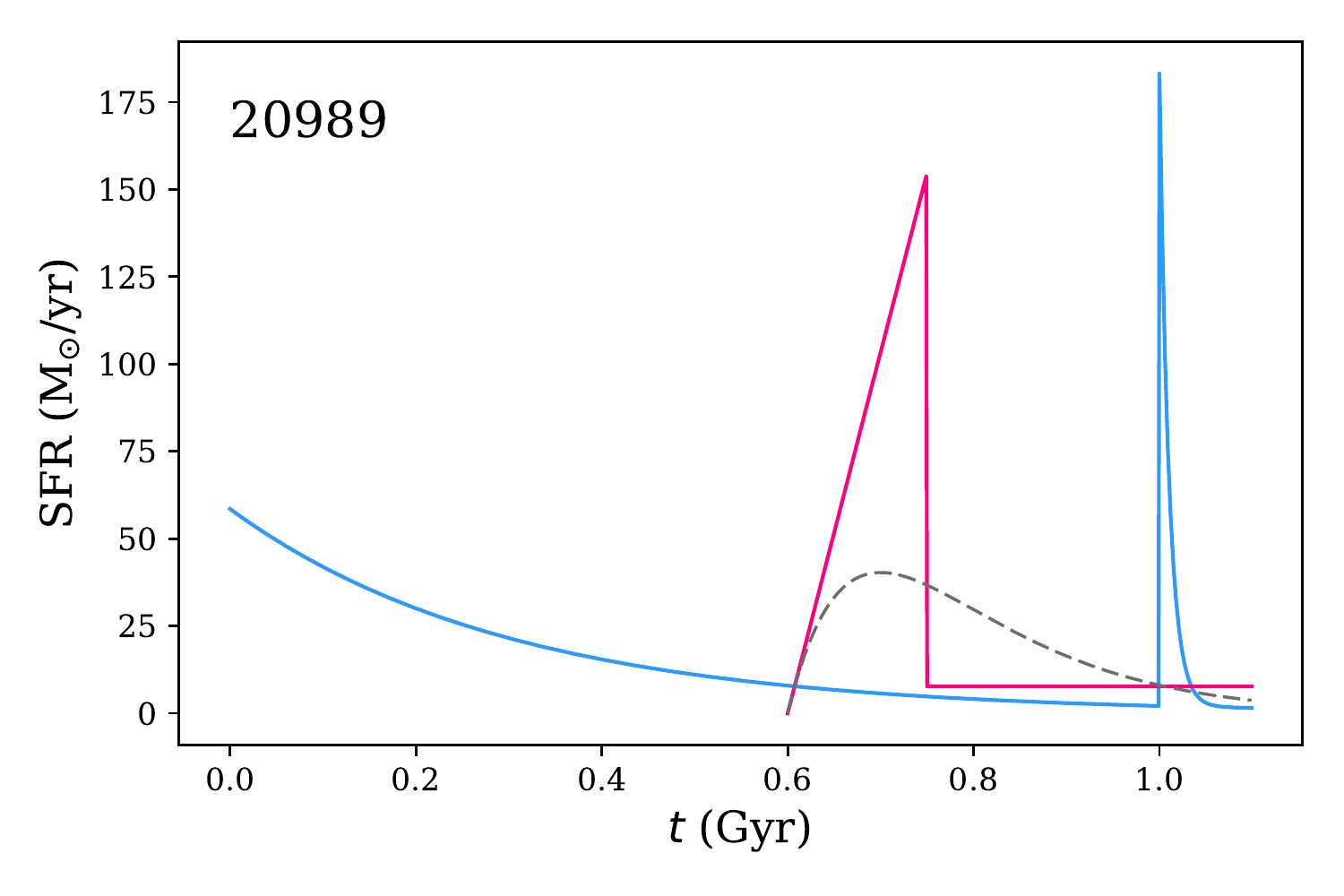}
	\includegraphics[width=6cm]{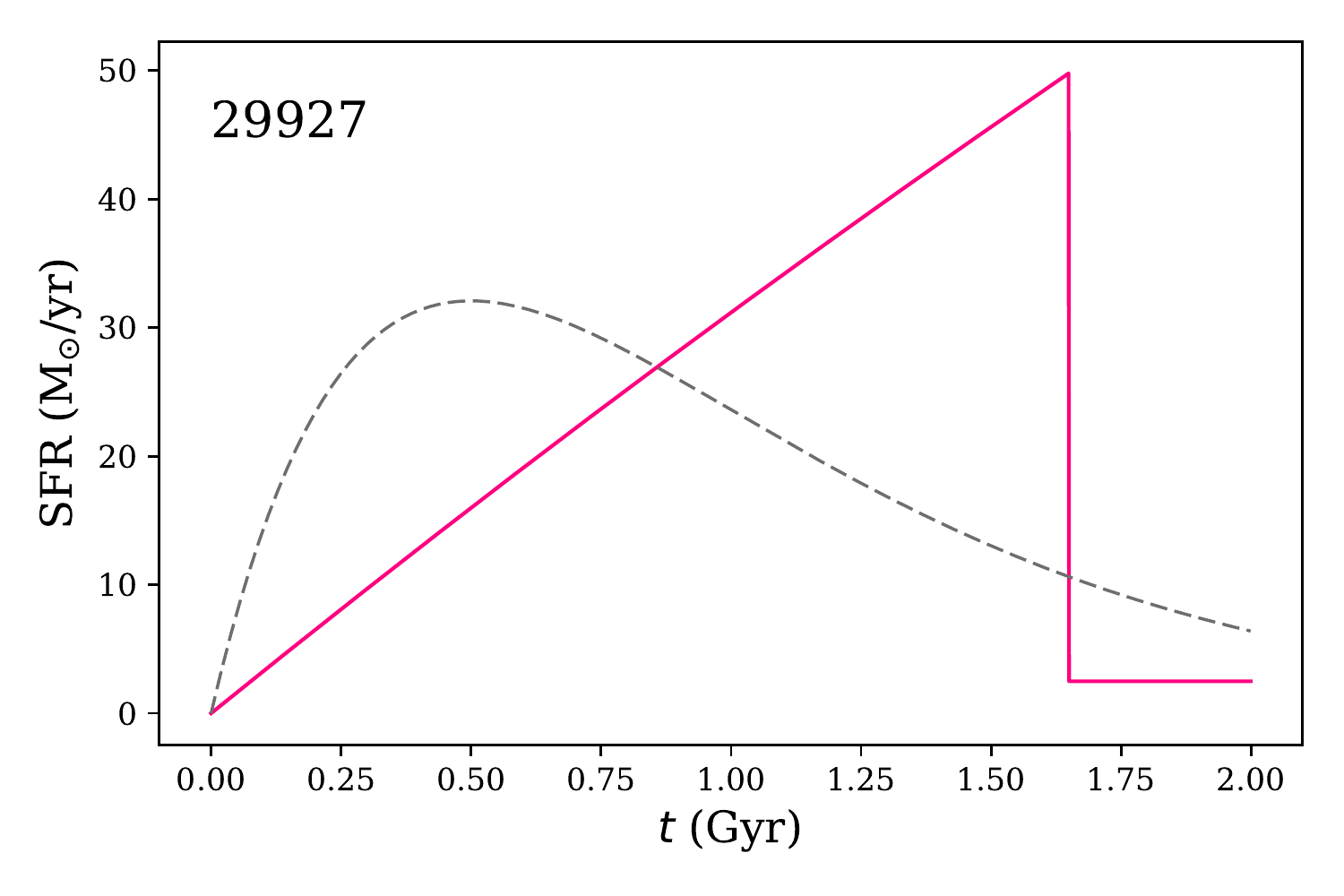}\\
 }
  	\caption{ \label{seds} Best fits and best SFHs obtained for 6 galaxies as example. The black points show the observations, the pink solid line the best fit obtained with the truncated SFH, and the gray dashed line the best fit obtained with the normal delayed-$\tau$ SFH. The six top panels show the SED and SFH of sources for which the two SFHs provide equivalent fit. The six bottom panels show examples of sources for which the delayed-$\tau$ SFH does not provide a satisfactory fit. The galaxies \#20946, \#20989, and \#29927 have been selected as candidates. For \#20989, the light blue curves in both the SED and SFH panels indicate the results obtained using a two populations SFH model (two exponentially-decreasing SFHs).}
\end{figure*}

The best fits of six randomly selected sources, three normal and three candidate galaxies having experienced a rapid downfall of SFR, are presented in Fig.~\ref{seds}.
The three normal galaxies are displayed in the top panels where we see that the best fit SEDs are identical whether the normal delayed-$\tau$ or the truncated SFH is used.
Their best fit SFHs are similar using either of the two SFHs, except for \#562 where the truncated SFH prefers the contribution of a recent burst to reproduce the SED whereas the delayed-$\tau$ shows a younger age to reach the final SFR.
The three other SEDs provide examples of fits results for three candidates.
For these galaxies, the delayed-$\tau$ SFH does not manage to fit both the UV and the NIR or reproduce the Balmer break.
Models built from a truncated SFH manage to reconcile the UV and NIR emission and provide good fit to the observed fluxes. 
For the three candidates the best truncated SFH clearly shows a rapid and drastic decrease.
For the delayed-$\tau$ SFH, in order to provide a equivalent final SFR, the age is shortened or the main peak of SFR is weaken.

\subsection{\label{testcase}The case study of three candidates}

\begin{figure*}
  	\includegraphics[width=6cm]{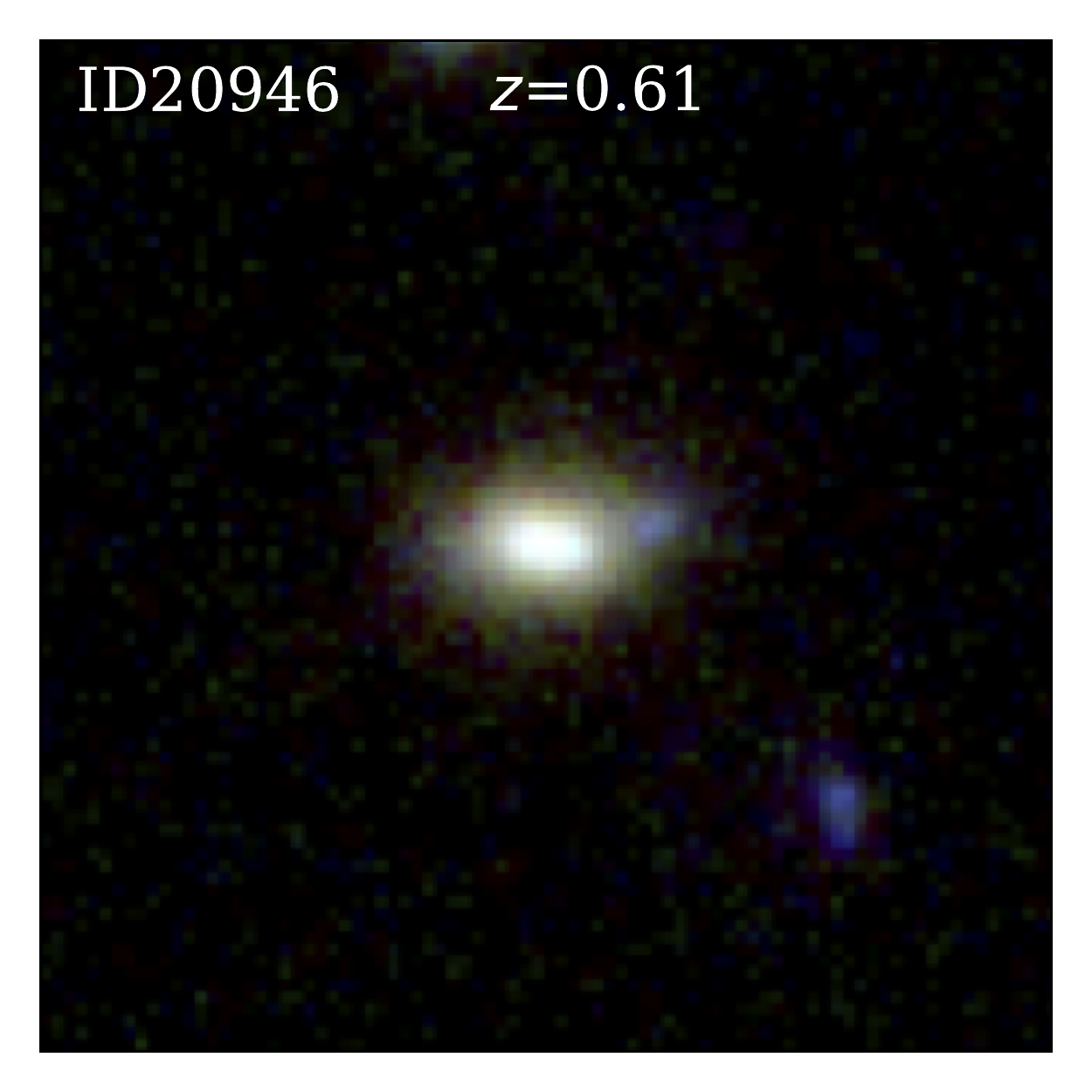}
  	\includegraphics[width=6cm]{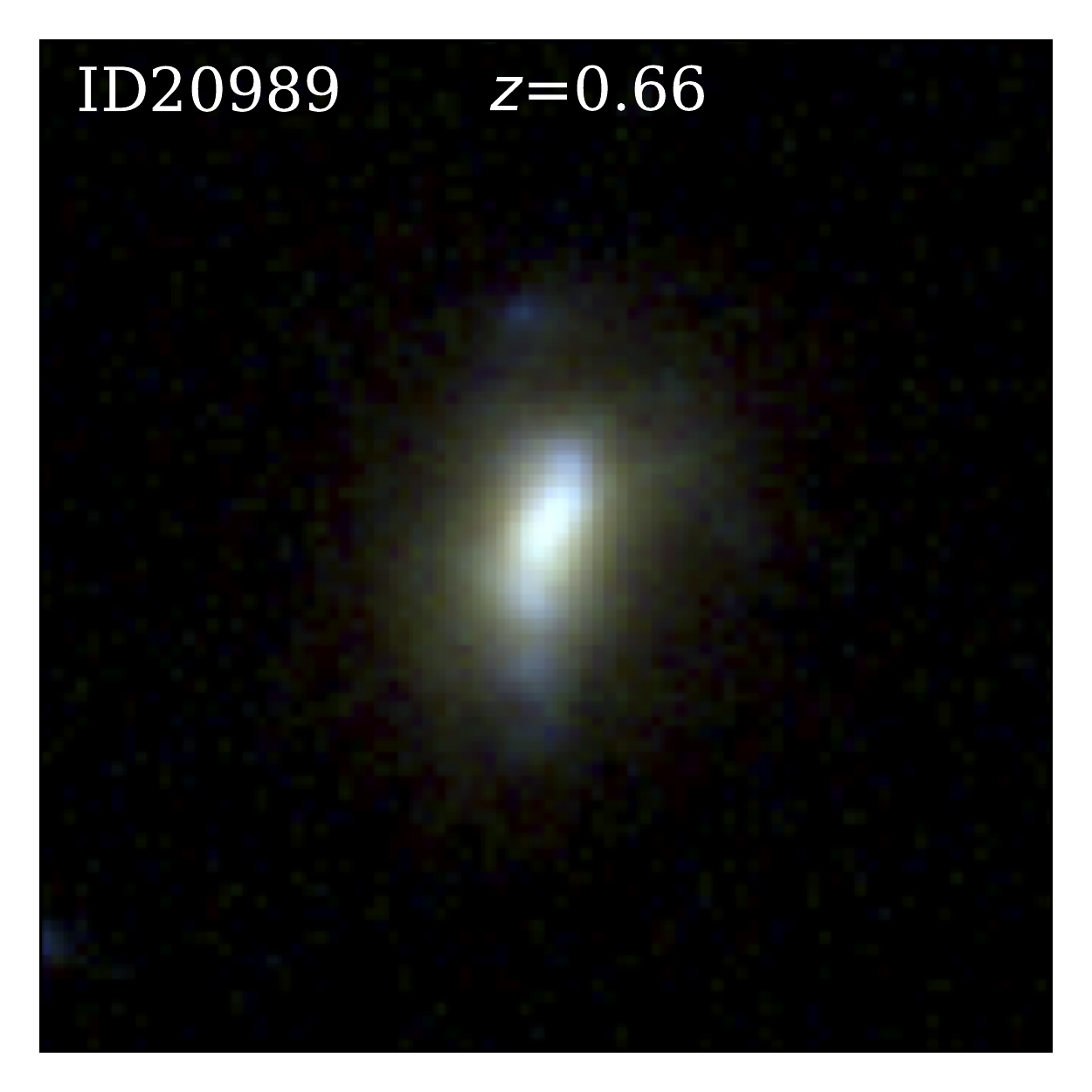}
  	\includegraphics[width=6cm]{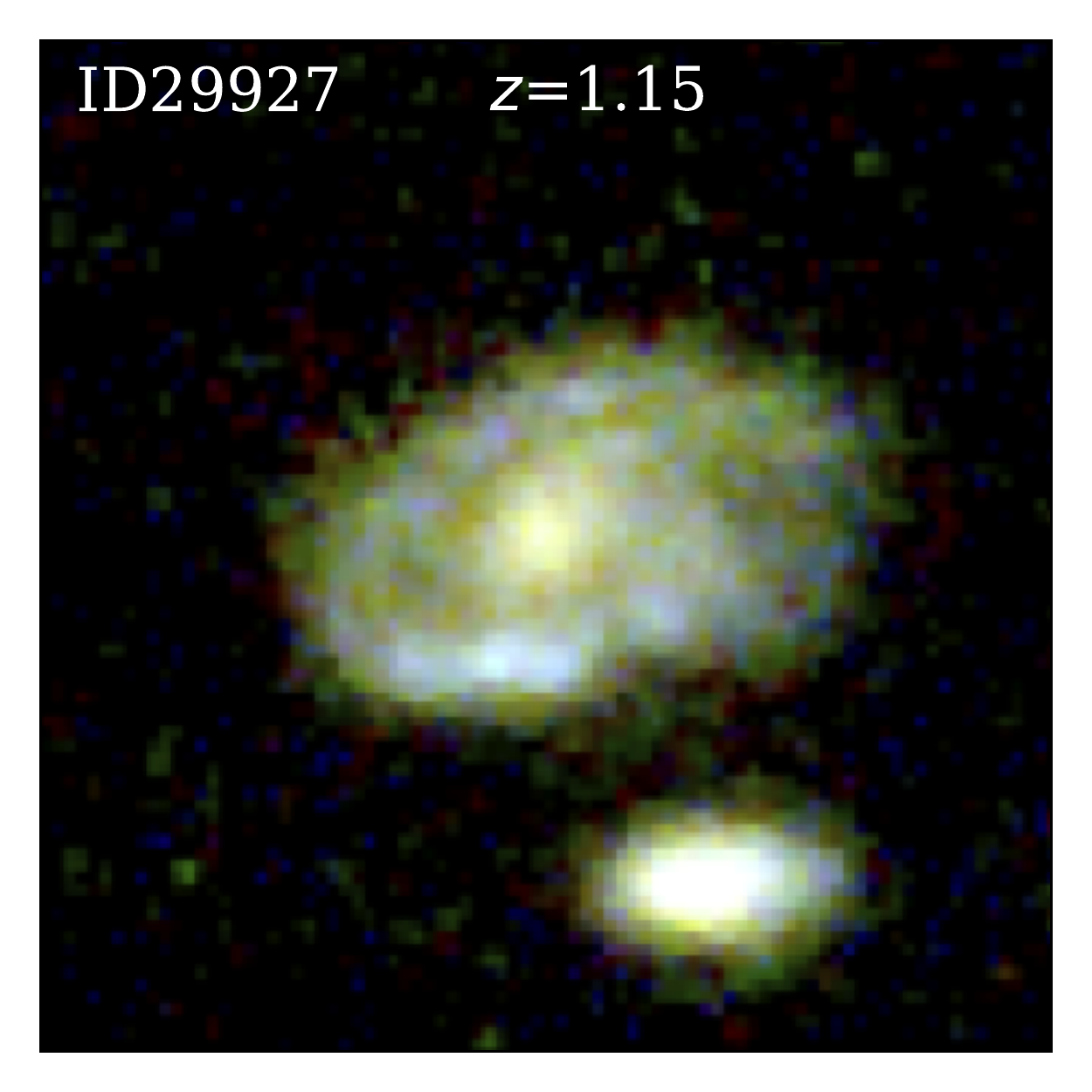}
  	\caption{ \label{rgb} RGB colors of the three candidates studied in details which SEDs are shown in Fig.~\ref{seds}.}
\end{figure*}

To understand the lower quality of delayed-$\tau$ SFH fits of compared to the truncated SFH, we choose to study in detail the three candidates, \#20946, \#20989, and \#29927, which SEDs are shown in Fig.~\ref{seds} and RGB images in Fig.~\ref{rgb}.
When running CIGALE, we are limited in the number of models to be fitted, and thus on the number of priors, due to computational reasons.
The size of the prior grids, if reasonable, does not affect our results on the estimate of the physical parameters since we derive them through their probability distribution function (if the minimum and maximum values are well-chosen and the grid contains a significant number of values).
However, it affects the value of the $\chi^2$ and thus can impact our selection criterion as the $\Delta$BIC is estimated using the $\chi^2$ of the fit for each model.
Thus, we check if $\Delta$BIC would be the same for these three galaxies if we used an increased grid of $\tau$, age, and E(B-V), and if they would still fulfill the criterion.
We run CIGALE using 100 values of $\tau$ (from 0.1 to 20\,Gyr), of age (from 1\,Myr to 13\,Gyr), and of E(B-V) (from 0.01 to 1.5), successively.
For each test, we provide in Table~\ref{tests} the resulting $\chi^2$ and $\Delta$BIC for both the truncated and normal delayed-$\tau$ SFHs.
In each of these tests, the $\chi^2$ obtained with the truncated SFH is the same with variation of less than 5\%.
For the normal delayed-$\tau$ SFH, the $\chi^2$ is improved up to 64\% in the case of an increased age grid for \#29927.
However, despite these improvements in the normal delayed-$\tau$ $\chi^2$, these galaxies still fulfill the selection criterion with a $\Delta$BIC larger than 10.
Thus the difference we observe in the quality of the fits using the two SFH is not due to the prior lists that we use for our main run.

		\begin{table*}
			\centering
			\caption{Robustness of the criterion against SED fitting priors.}
			\begin{tabular}{l c c c c c c c c c c c c c c }
	 		\hline\hline
			id & n$_{bands}$ & \multicolumn{2}{c}{main run} 	    & 	 \multicolumn{2}{c}{increased grid of $\tau$} & 	 \multicolumn{2}{c}{increased grid of age} & \multicolumn{2}{c}{increased grid of E(B-V)} & \multicolumn{2}{c}{UV no fitted}  	\\
			    &                          & $\chi^2_{del}$ & $\chi^2_{trunc}$ & 	$\chi^2_{del}$ & $\chi^2_{trunc}$                    & 	$\chi^2_{del}$ & $\chi^2_{trunc}$                & 	$\chi^2_{del}$ & $\chi^2_{trunc}$  & 	$\chi^2_{del}$ & $\chi^2_{trunc}$\\ 
			    &                          & \multicolumn{2}{c}{$\Delta$BIC}    &	 \multicolumn{2}{c}{$\Delta$BIC}                      &	 \multicolumn{2}{c}{$\Delta$BIC}                  & 	 \multicolumn{2}{c}{$\Delta$BIC}    & 	 \multicolumn{2}{c}{$\Delta$BIC}\\ 
			\hline
			\#20946   & 38 & 97.9 & 67.2					    &	             97.9              &            67.2                                       &  83.5                            &     65.8            &      91.1                          &       64.5       &         5.4             &             3.9               \\
			                &       & \multicolumn{2}{c}{27.1}  			    &	\multicolumn{2}{c}{27.1}                                             &	\multicolumn{2}{c}{14.0}                                          &	\multicolumn{2}{c}{23.0}   &	\multicolumn{2}{c}{-2.2}    \\
			\hline
			\#20989 & 36& 107.0 & 	65.2				    &	         107.0                  &      63.2                                            & 83.5                             &       63.0                                       &        80.3                        &        62.6  &           10.8                     &        9.9                    \\
			                &       &\multicolumn{2}{c}{38.2} 			    &	\multicolumn{2}{c}{40.2}                                             &	\multicolumn{2}{c}{16.9}                                           &	\multicolumn{2}{c}{14.1}    &	\multicolumn{2}{c}{-2.3}\\
			\hline
			\#29927 & 32&  74.7 & 32.2					    &	          74.5                 &       32.2                                           &  45.6                            &       29.1                                       &         58.8                       &         32.2   &         5.6                  &   4.8                         \\
			                &      & \multicolumn{2}{c}{39.0}			    &	\multicolumn{2}{c}{38.8}                                              &	\multicolumn{2}{c}{13.0}                                          &	\multicolumn{2}{c}{23.0}    &	\multicolumn{2}{c}{-2.3}\\
			\hline
			\label{tests}
			\end{tabular}
		\end{table*}	

The UV should be the first spectral range to be impacted by recent variations in the SFH.
Thus we test if fitting the SED of our three test galaxies without the UV rest-frame ($<$300\,nm) fluxes would lead to a better agreement between the two SFHs.
The runs are made with the increased range of age priors as we just showed that it improves the $\chi^2$ obtained with the normal delayed-$\tau$ SFH.
The $\Delta$BIC obtained for these runs are negative with an absolute value lower than 10 indicating that both SFH provide an equivalent fit of the data (Table~\ref{tests}).
The UV is thus the key domain that the delayed-$\tau$ SFH struggles to reconcile with the optical-NIR part of the SED.
However, as shown in Fig.~\ref{seds}, both the delayed-$\tau$ and truncated SFHs provide a good fit of the UV emission of \#20989.
Here, the discrepancy comes from the optical/NIR emission.
One can then imagine that with the addition of an older population, that is considering two stellar populations, it would be possible to fit the NIR emission without the need for a recent fast decrease of the SFR.
To test this possibility, we fit the SED of \#20989 using a double exponentially-declining SFH, available in CIGALE, that allows that mimics two populations combined through an extra free parameter.
The $\chi^2$ obtained with this SFH is 82.4 versus 65.2 for the truncated SFH although the two exponentially-declining SFH has more free parameters.
Thus a more complicated SFH is not preferred in this case.
Interestingly, as shown in Fig.~\ref{seds} the double exponentially-declining SFH favors a main stellar population with quasi constant evolution plus a recent burst of SF followed by a rapid decline.
The same test was applied to all the candidates and none of them was significantly better fit by a two population SFH.
Thus, we conclude that the candidates underwent rapid and recent variations in their SF, probed by the UV, difficult to be modeled by a smooth SFH.

\begin{figure}
  	\includegraphics[width=\columnwidth]{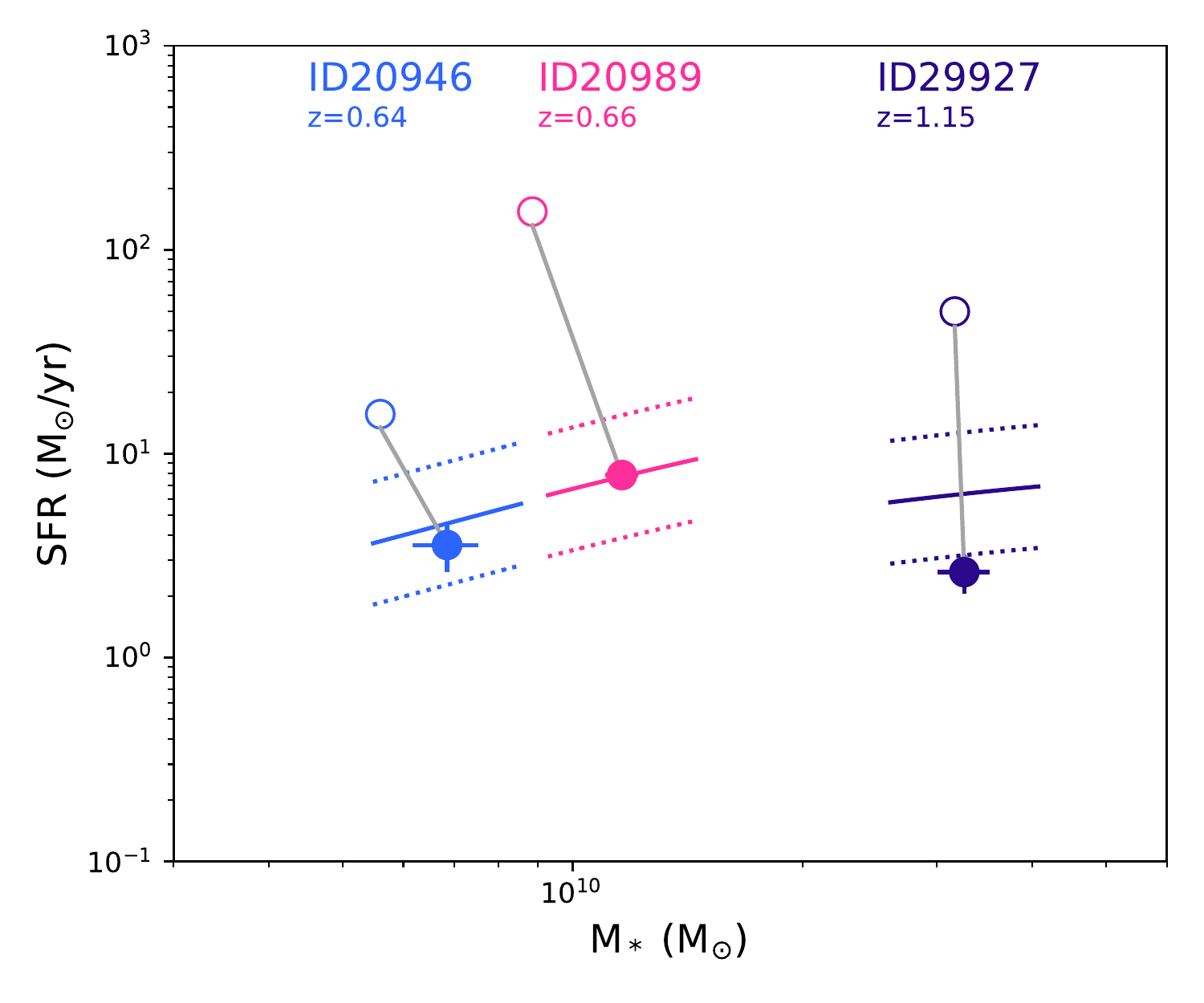}
  	\caption{ \label{ms_cand} Positions of the three test candidates relative to the MS after (filled circles) and before (empty circles) the drastic decrease of star formation activity. The MS and its dispersion is indicated at the redshift of each candidate with a solid and dashed lines, respectively.}
\end{figure}

We show RGB color images of the three test candidates in Fig.~\ref{rgb}.
The sources \#20946 ($z$=0.61) and \#20989 ($z$=0.66) have a red disk and exhibit some UV emission as lanes.
Their morphology suggest a merger scenario.
In Fig.~\ref{ms_cand}, we place them on the SFR-M$_*$ plane along with the MS at their respective redshift.
Both \#20946 and \#20989 lied above the MS prior to their decrease of SFR.
Using the information provided by the SED fitting on the strength of the SFR decrease and knowing that it happened approximately in less than 500\,Myr (the exact time is not constrained from broad-band SED fitting, see \cite{Ciesla16} and \cite{Boselli16b}), we can recover their position on this plane before the downfall of SFR.
They were thus previously even higher above the MS by a factor of 6.7 and 33.3  for \#20946 and \#20989, respectively, comforting the merger scenario.
The question is then to know whether they will later remain on the MS or continue to fall and become passive.
Fast decreases of SFR of these orders are consistent with the simulated SFH of mergers computed by \cite{DiMatteo05}.
Interestingly, even the morphologies of \#20946 and \#20989 with a red disk and UV lane across the disk are consistent with the snapshots of their simulation.
Since \#20946 and \#20989 are relatively low mass galaxies, below 1.2$\times$10$^{10}$\,M$_{\odot}$, we do not expect these mergers to have a strong dust enshrouded phase, allowing us to identify them from UV-to-NIR broadband SED modeling.
The case of \#29927 is slightly different as the galaxy is now below the MS.
Previously it was indeed above, but, went down in a few hundreds Myr.
This source is more massive with $\log$M$_*$=10.51 and at higher redshift ($z$=1.15).
Its morphology (Fig.~\ref{rgb}, right panel) shows a large red disk with a central bulge and a UV ring in the middle of the disk.
There is also another source lying close to \#29927 that could be an indication of possible interactions.

\section{\label{properties}Physical properties of the candidate galaxies }

\begin{figure*}
  		\includegraphics[width=\textwidth]{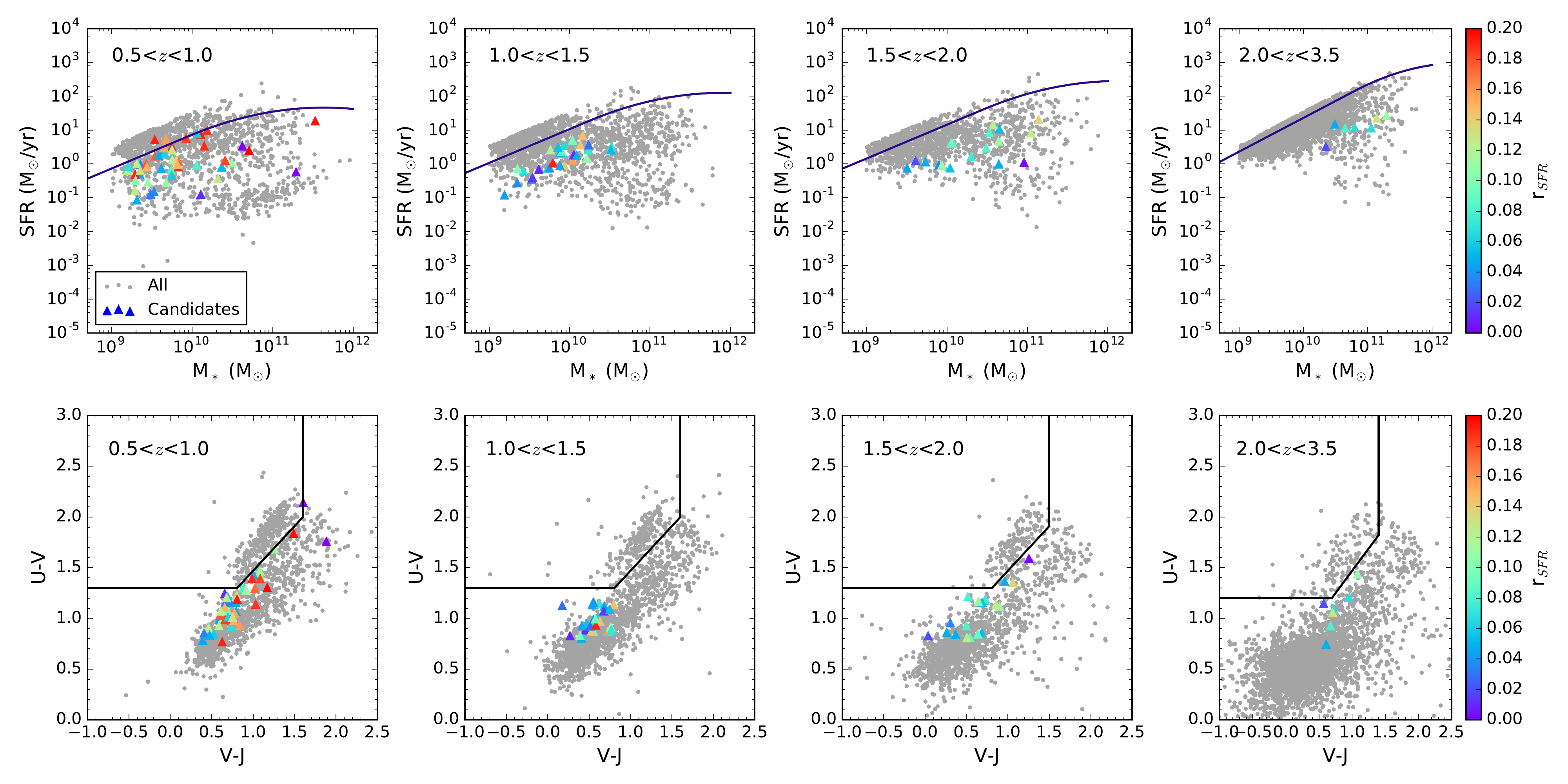}
  		\caption{ \label{msuvj} Top panels: Fitted galaxies placed on the main sequence diagram. The four panels show the main sequence in four redshift bins. Gray dots show the location of galaxies with good fits obtained with both SFH models, the pink triangles show the location of the galaxies selected as candidates for a possible recent drop of their star-formation activity. Blue solid lines indicate the position of the MS at the mean redshift of the bin \citep{Schreiber15}. Bottom panels: Galaxies displayed on the UVJ diagram. Grey dots are the galaxies of the sample well fitted by both SFH models. Colored dots are the sources selected, color-coded according to the $r_{\rm{SFR}}$ parameter. The black solid lines indicate the UVJ selection according to \cite{Whitaker11}.}
	\end{figure*}

	To investigate the properties of these 102 sources, we place them on the main sequence diagram (Fig.~\ref{msuvj}).
	They are mostly located in the lower part of the MS indicating a possible transition in their star formation activity.
	This is consistent with the hypothesis that these sources, selected from their SED, are undergoing a diminution of their SFR.
	Moreover, we observe that the candidates span a large range of stellar masses, from $\sim$10$^9$\,M$_{\odot}$ to masses higher than 10$^{11}$\,M$_{\odot}$.
	Adding the information on the $r_{\rm{SFR}}$ parameter allows us to see that the candidates located in the lower part of the MS still have some residual star formation showing that they are not totally quenched.
	Indeed, our criterion imposes a drastic diminution of about 80$\%$ of their star formation activity in the last few hundred Myr and not necessarily a total quenching.
	We then display the candidate sources on the UVJ diagram.
	We recall that our candidates are pre-selected to lie in the star-forming region of this diagram, which explains why none of them lies in the quiescent region. 
	However, they are located in a region just below the passive region, close to the line separating the two zones, supporting the hypothesis that these galaxies are undergoing or just underwent a decrease of star formation activity.

\begin{figure}
  		\includegraphics[width=\columnwidth]{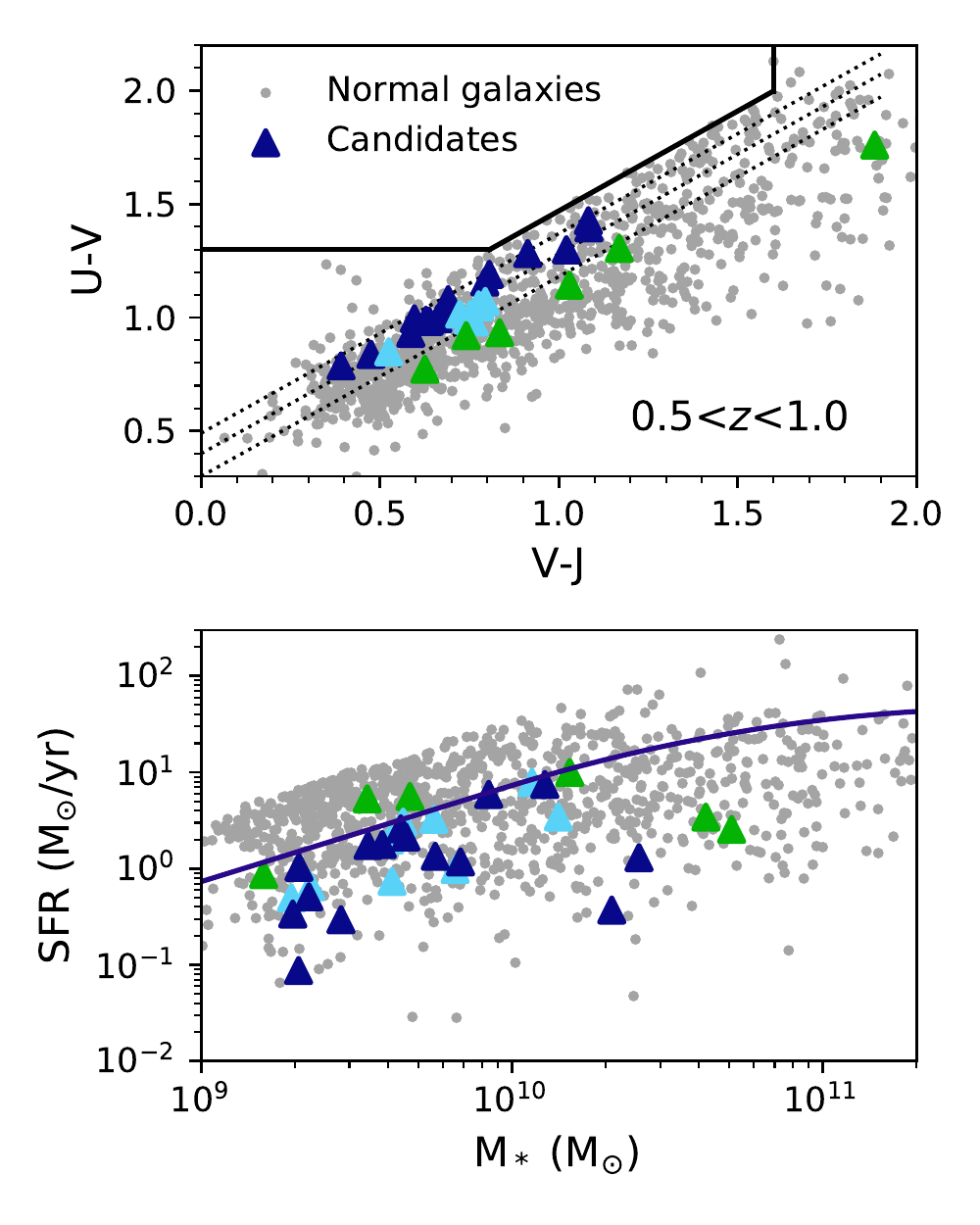}
  		\caption{ \label{msuvjslices} UVJ and MS diagrams for the first redshift bin (0.5$<z<$1.0). The normal galaxies are shown in gray whereas the candidates (crosses) are separated in slices in the UVJ diagram. The slices used to separate the candidates are indicated with the black dotted lines.}
	\end{figure}

	Recent studies have shown that galaxies displayed in color diagrams, such as UVJ or NUVrK, show a gradient of sSFR decreasing from the active region toward the passive region \citep{Arnouts13}.
	We thus check if our selection is equivalent to a sSFR selection.
	To test this point, we separate our selected sample in three slices in the UVJ diagram, parallel to the line separating the active and passive region (Fig.~\ref{msuvjslices}, top panel).
	If they were selected from their sSFR value, we would expect that these three slices will be recovered in the MS diagram.
	But, as shown in Fig.~\ref{msuvjslices} (bottom panel), we recover no gradient of sSFR in the MS diagram, the galaxies are dispersed.
	This means that our criterion is not just a selection in sSFR and that we are able to select galaxies with different SFH than the galaxies located in the same zone in the UVJ diagram.
	Indeed, as discussed for the three test sources in Sect.~\ref{testcase}, the UV is a key domain to probe recent variations in the SFH and ``normal'' analytical SFHs fail to reconcile the UV and NIR colors of our candidates.
	But the UV is not used in the UVJ diagram explaining why we can find galaxies with identical UVJ colors but different SFHs.

	\subsection{Reference sample}
	
	The position of the candidate galaxies on the UVJ diagram show that they are at the limit between the passive and the active galaxies regions that could imply that these candidates may be undergoing some SFR downfall.
	Yet, other galaxies are located in the same regions of the UVJ diagram and do not show any specific SED shape.
	To understand the difference between the candidates and the others, we define a reference sample.
	For each candidate, we select the galaxies that are located in the UVJ diagram at the same position and in the same redshift bin, within a circle defined as:
	\begin{equation}
		\sqrt{\left [(V-J) - (V-J)_{q}\right]^2 + \left[(U-V) - (U-V)_{q}\right]^2} \leq 0.2,
	\end{equation}
	\noindent where $(V-J)$ and $(U-V)$ are the colours of the reference galaxies and $(V-J)_{q}$ and $(U-V)_{q}$ those of the candidate.
	The 0.2 value is chosen arbitrarily after some tuning to ensure enough statistics to compute average properties of the reference sample.
	Tests have been made to ensure that the results of this study do not depend on reasonable variation of the adopted radius.

	\subsection{Physical parameters}
	
\begin{figure*}
  		\includegraphics[width=\textwidth]{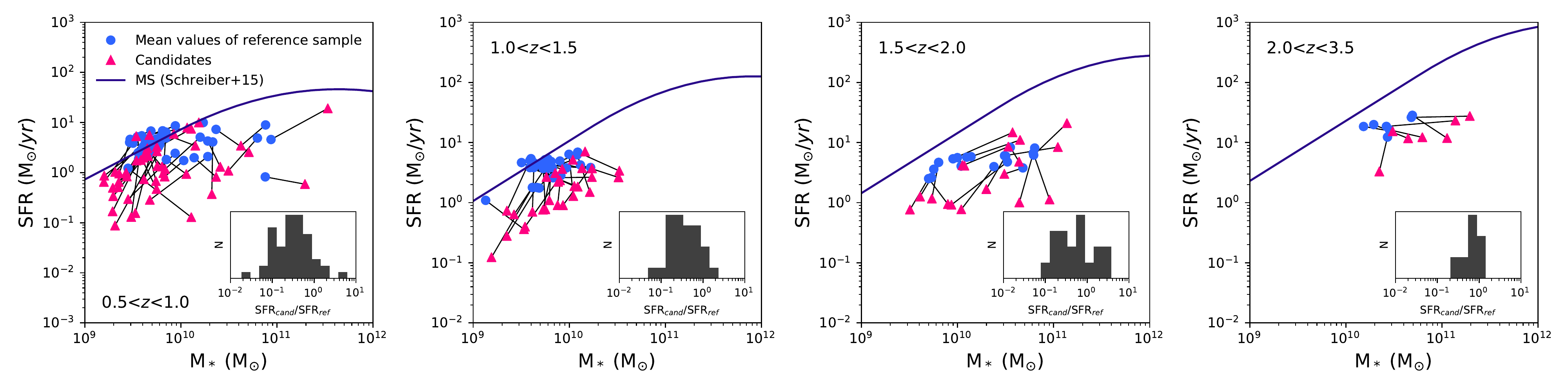}
  		\caption{ \label{msssfr} Comparison between the positions of the candidate sources (red triangles) and the mean values of their reference sample (blue circles) in the SFR--Mass diagram.  The inset panels show the distribution of the ratio between the SFR of the candidate galaxies and their respective sample SFR.}
	\end{figure*}

	We compare the physical properties of the candidates to those of the reference sample.
	We average the properties of the galaxies of each reference sample and compare the results to those of the corresponding candidate.
	In Fig.~\ref{msssfr}, we place on the SFR-M$_*$ plane the mean values of the stellar masses and SFR of the reference sample corresponding to each candidates in the four redshift bins.
	The position of each reference sample, determined from its average properties, is systematically different than the position of the corresponding candidate galaxy.
	As expected, the SFR of the candidates is systematically lower than the average SFR of normal galaxies located in the same place of the UVJ diagram.
	
\section{\label{morphology}Morphology}
	
With our method, we have adopted an empirical criterion to select galaxies that recently underwent a drastic decrease of their SFR and showed that their physical properties are different from those of galaxies with identical (U-V) and (V-J) colors.
We now look at the morphology of the candidates.
The morphological classification of \cite{HuertasCompany16} using deep machine learning is used to separate the candidates between spheroids (24 galaxies), pure disks (31 galaxies), disk plus spheroids (4 galaxies), irregular disks (20 galaxies), and mergers or irregular systems (12 galaxies).

\begin{figure*}
  	\includegraphics[width=\textwidth]{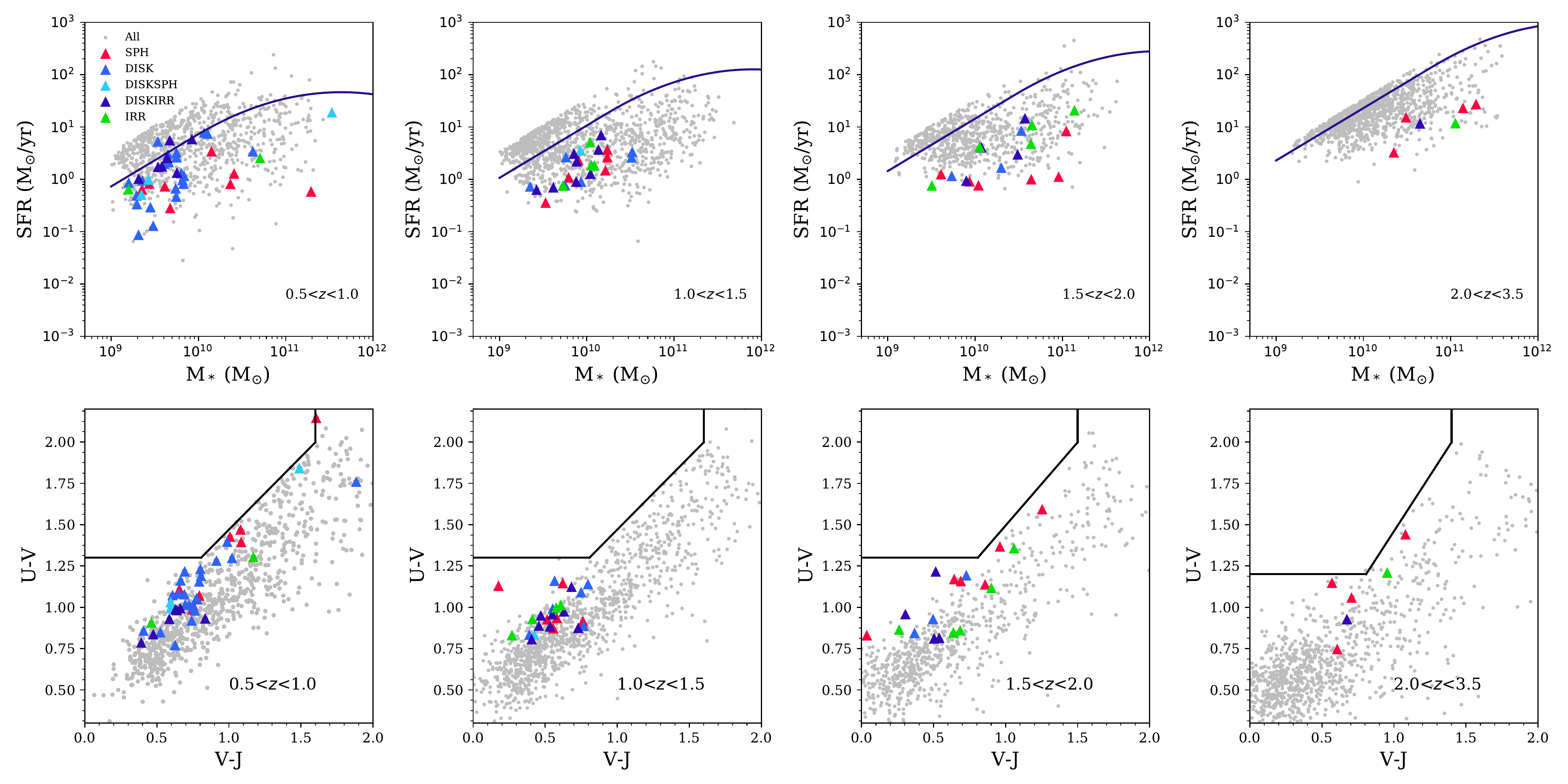}
  	\caption{ \label{uvj_morpho} Main sequence (top row) and UVJ (bottom row) diagrams. The candidates have been color-coded according to their morphology classification as defined in \cite{HuertasCompany16}: in pink the spheroids, in blue the disks, in orange the disk+spheroids, and in green the irregular galaxies.}
\end{figure*}

\begin{figure}
  	\includegraphics[width=\columnwidth]{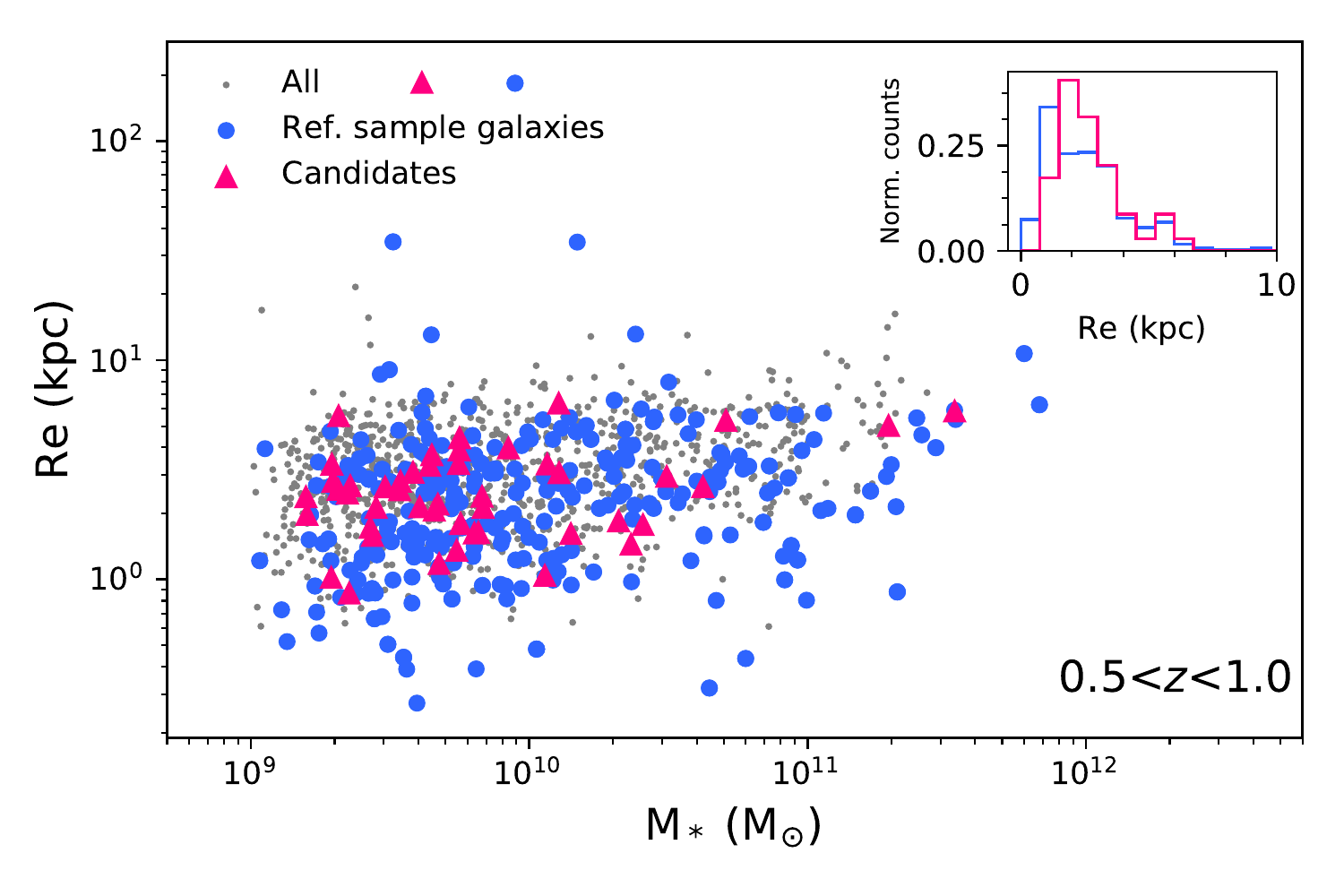}
  	\caption{ \label{re} Effective radius of galaxies with redshift between 0.5 and 1 as a function of their stellar mass. Grey dots are all the galaxies in this redshift range. Pink triangles are the candidates of this redshift range, and the blue circle are all the galaxies lying in the same part of the SFR-M$_*$ plane than the candidates, that is the galaxies used to build the reference samples. The inset panel shows the distributions of $R_e$ in kpc for both the candidates and the galaxies associated to the reference samples.}
\end{figure}

We show in Fig.~\ref{uvj_morpho} the candidates in the MS and the UVJ plots, color coded as a function of their morphological classification.
There is no segregation of one particular type in a specific stellar mass range, spheroids candidates are found at both low and high masses as well as disks.
However, at $z$$>$2, the candidates are almost all  spheroids or mergers, with only three disks out of 24 galaxies.
Comparing the morphological types of the candidates with those of a reference sample is not straightforward. 
Instead, we look at their effective radius, $R_e$ computed by \cite{VanderWel14}.
As shown in Fig.~\ref{re}, the candidates and the galaxies of their reference samples cover the same range of effective radius (see inset panel of Fig.~\ref{re}).
The median effective radius are 2.3\,kpc and 2.5\,kpc for the reference samples galaxies and the candidates, respectively.
The first and third quartiles of both distribution are also consistent: 1.3 and 3.3\,kpc for the reference samples galaxies and 1.8 and 3.4\,kpc for the candidates.
Therefore, we conclude that they have effective radius that are consistent with those of galaxies of similar stellar masses.

\section{\label{agn}AGN activity}
		
A possible source of rapid downfall of star-formation could come from AGN activity as some simulations predict the AGN-driven outflows to be responsible for the downfall of star formation activity \citep[e.g.,][]{Baldry04,Hopkins06} although other results suggest than these outflows hardly affect the gas content of galaxies \citep{Gabor14}.
Out of the 102 candidate galaxies, 5 are detected in X-ray \citep{Cappelluti16} and only 4 of them have an X-ray luminosity above 10$^{42}$\,erg/s: \#1817, \#3463, \#10032, and \#30250.
The 4 sources with L$_X$$>$10$^{42}$\,erg/s are massive with a stellar mass equal or higher than 1.7$\times$10$^{10}$\,M$_{\odot}$.

Furthermore, we test whether some of these sources could host an enshrouded AGN emitting in IR.
This can only be explored for sources detected in the MIR/FIR, we thus select the 47 candidate sources detected at MIPS 24\microns.
Using the AGN emission modeling of the CIGALE code, we estimate for each source the need for an AGN template to fit the UV to IR SED.
Out of the 47 sources, 23 have an output $frac_{\mathrm{AGN}}$ parameter (contribution of the AGN to the total IR luminosity) below 10\%.
Based on the results of \cite{Ciesla15}, we can conclude that these sources do not likely host a strong IR AGN.
Twenty-three sources have a trustable $frac_{\mathrm{AGN}}$ parameter above 15\%.
However, among these 23 likely-AGN-galaxies, 18 of them have only one IR flux on which the decomposition can rely, which is not enough to provide a fiducial conclusion on the presence of an enshrouded AGN in these sources.
Five candidates, however, have a high $frac_{\mathrm{AGN}}$  between 30 and 50\%: \#1415, \#10534, \#15220, \#17144, and \#18949.
Out of these five sources, the SED of four of them indicates an unusually high IR 24\microns\ or 100\microns\ flux compared to the expected ones from the energy absorbed in UV-NIR (CIGALE performs energy balance), suggesting a possible blending problem in the IR photometry.
This is confirmed by the fact that all of these four sources have a close companion. 
Only \#18949 shows a clear sign of possible AGN contribution in MIR.
	
Out of the 102 candidates, only 5 of them show a sign of AGN activity either in X-ray or in IR.
If it was to be interpreted in terms of duty cycle, considering that our method is sensitive to variations occurring less than 500\,Myr prior to observation, it would imply an AGN duty cycle of 5$\%$.
Assuming that all the candidates went through an AGN-on phase, it would lead to an AGN visibility timescale of 25\,Myr. 
However, given the poor statistic of AGN detected candidates, we conclude that AGN might not be the main cause for the recent star-formation downfall underwent by the candidate sources.

\section{\label{variations}Probing recent movements relative to the main sequence}

\begin{figure}
  	\includegraphics[width=\columnwidth]{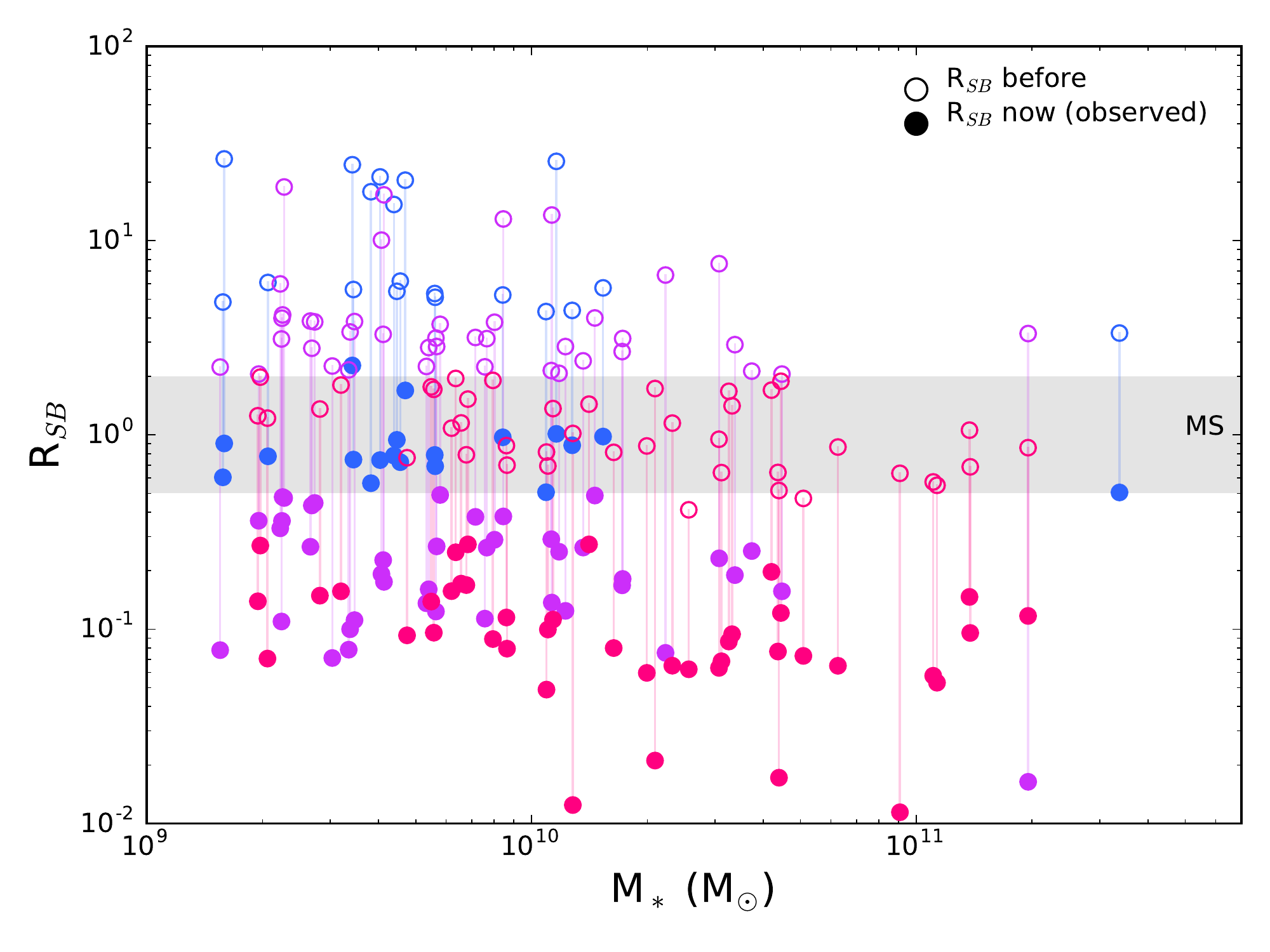}
  	\caption{ \label{rsb} Starburstiness of the candidates before and after the strong decrease of their star-formation activity as a fonction of their stellar mass. Filled circles indicate the position of the candidate at the moment there are observed, and the empty circles show the position of the candidates prior the SF decrease. Colors indicate different movements: from the starburst region to the MS (blue), from above to below the MS (purple), and from the MS to below the MS (red). The grey regions show the MS and its scatter.}
\end{figure}

\begin{table}
	\centering
	\caption{Fractions of spheroids, disks, and irregular galaxies for candidate and reference galaxies, respectively.}
	\begin{tabular}{l c c}
	 \hline\hline
	Morphology & Candidates & Reference galaxies \\ 
	\hline\hline
	\multicolumn{3}{c}{First group (postSB galaxies)}\\  
	\hline
	Spheroids	& 0	\%& 23 \%\\
	Disks	& 56	\%& 20 \%\\
	Irregulars	& 44	\%& 57 \%\\
	
	\hline
	\multicolumn{3}{c}{Second group (from above to below the MS)}\\  
	\hline
	Spheroids	& 30 \%& 24 \%\\
	Disks	& 30 \%& 21 \%\\
	Irregulars	& 40 \%& 55 \%\\
		
	\hline
	\multicolumn{3}{c}{Third group (from the MS to below the MS)}\\  
	\hline
	Spheroids	& 36 \%& 25 \%\\
	Disks	& 39 \%& 29 \%\\
	Irregulars	& 25 \%& 46 \%\\
		
	\label{mt}
	\end{tabular}
\end{table}

The epoch when the fast decrease of star formation occurred remains unconstrained from broadband photometry alone. 
However, the SFR that the candidate galaxies experienced before the decrease is constrained by the $r_{\rm{SFR}}$ parameter. 
This information allows to locate the position of the galaxies with respect to the star-formation MS prior to the downfall of SFR, i.e., to determine their past starburstiness ($R_{SB}$, see Fig.~\ref{rsb}). 
The candidates fall in three categories discussed below.

\begin{figure*}
  	\includegraphics[width=\textwidth]{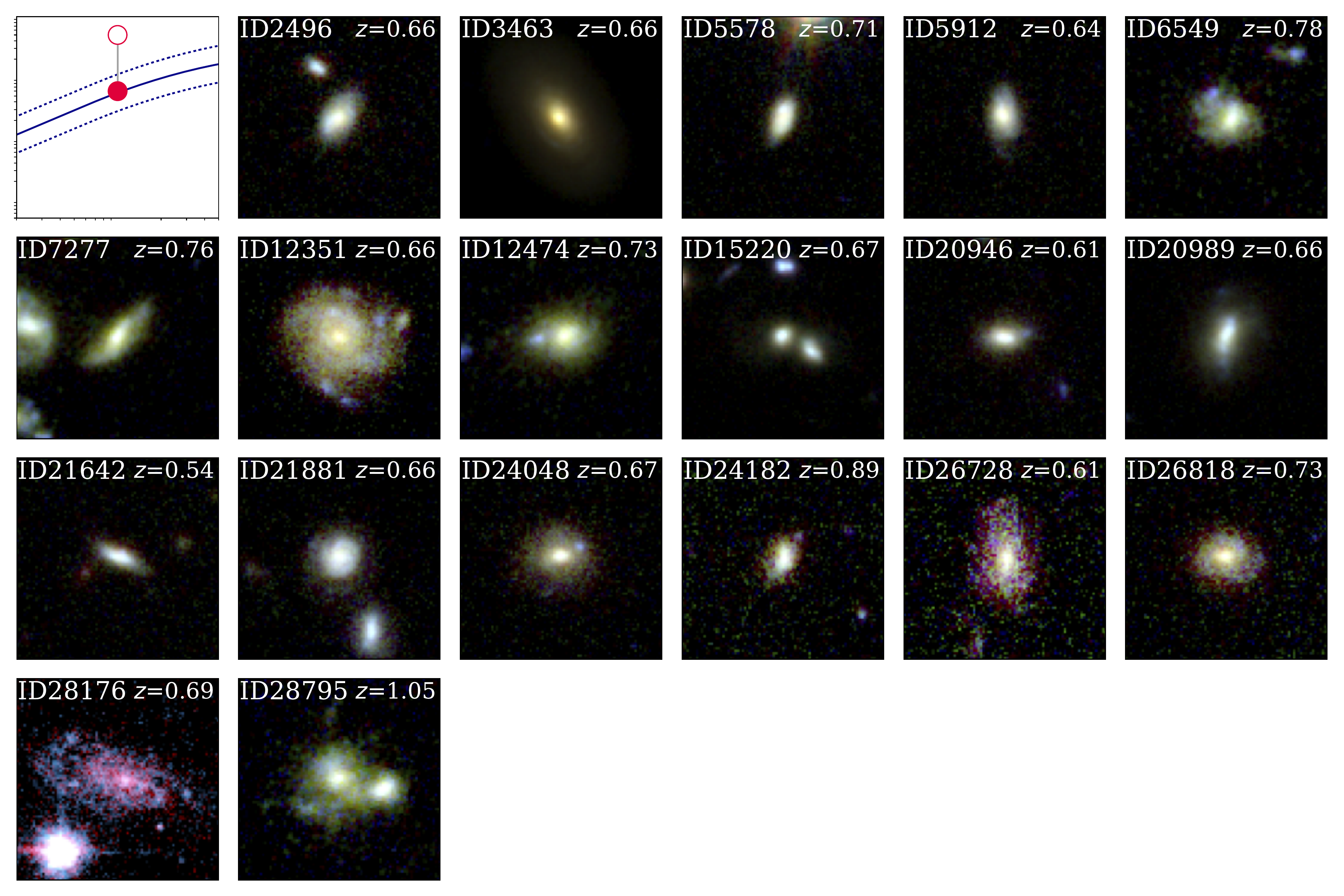}
  	\caption{ \label{image_mvm_0} Candidates that were lying in the starburst region before undergoing a fast downfall of star formation and that are now lying on the MS.}
\end{figure*}

The first group is composed of sources that were lying in the starburst region, i.e. above the MS, before undergoing the decrease of star formation activity and are now on the MS (Fig.~\ref{image_mvm_0}).
These sources are thus post-starburst candidates.
They have stellar masses lower than 2$\times$10$^{10}$\,M$_{\odot}$. 
Only one source is massive and is the most massive of the sample, \#3463.
According to the morphological classification of \cite{HuertasCompany16}, 56\% of the galaxies of this group are disks.
This is almost a factor three more than what is found for galaxies located in the same region of the SFR-M$_*$ (see Table~\ref{mt}).
The remaining 44\% are classified as irregular disks or systems versus 57\% found for ``normal galaxies''. 
We notice the absence of spheroids in this candidate group when we expect about 23\% of the ``normal'' galaxies to have this morphological type.
With the data in hand, we cannot state if these candidates are still in transition and thus just passing through the MS and will reach the passive region afterwards or if, after their star formation burst, they are back on the MS and will stay in this position for a given time.
Going further in this analysis would require, at least, spectral information on H$\alpha$ for instance, to compare different star formation indicators sensitive to different timescales.

\begin{figure*}
  	\includegraphics[width=\textwidth]{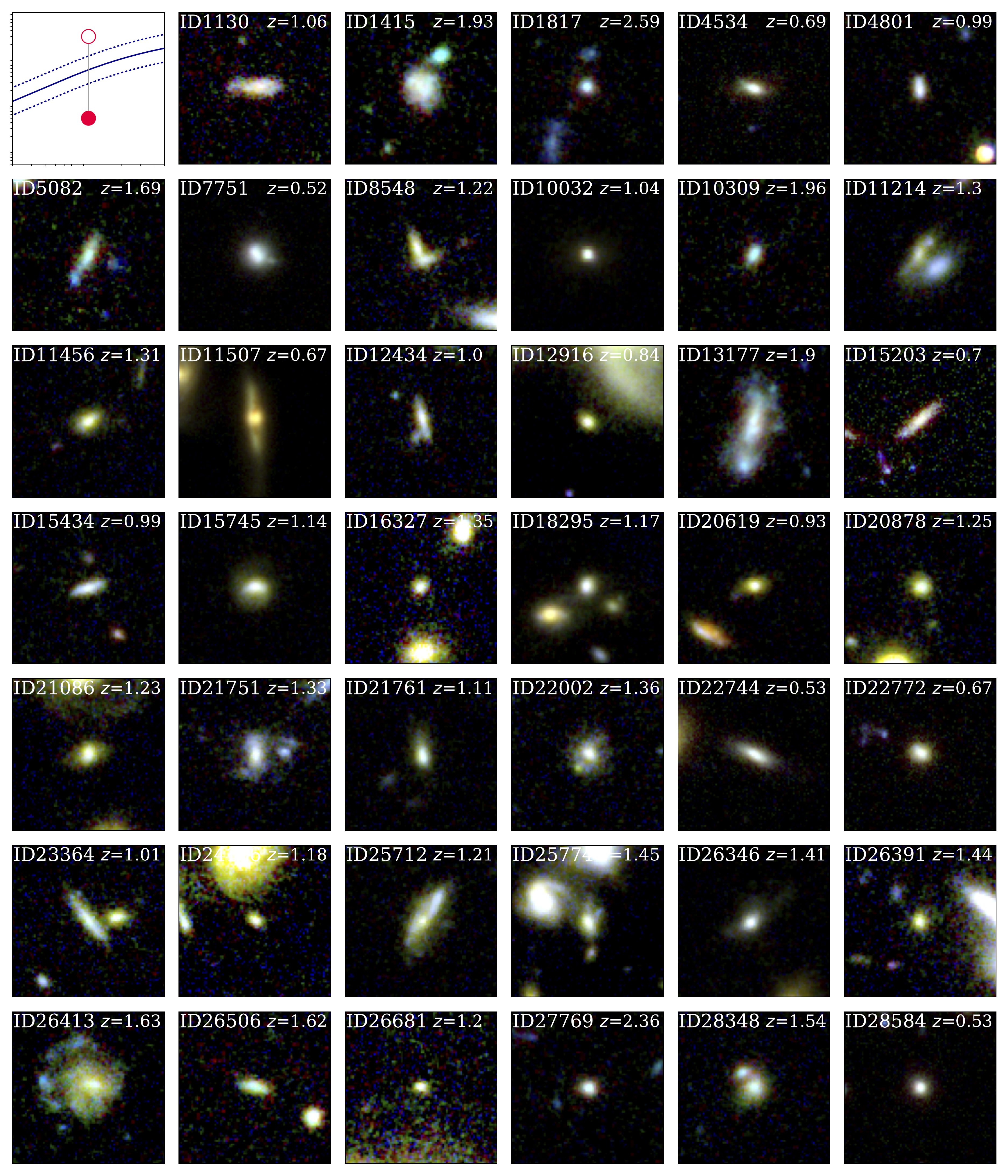}
  	\caption{ \label{image_mvm_1} Candidates that were in the starburst region before the slowdown of their star formation and that are now below the MS.}
\end{figure*}

The second group gathers candidates that were in the starburst region and that are now below the MS (Fig.~\ref{image_mvm_1}).
These sources, which have stellar masses spanning a larger range,  passed through the MS in a relative short time which is of the order of several hundred Myr.
The distribution in terms of morphological types in this group is more balanced as we find 30\% of spheroid galaxies, 30\% disks or disk+spheroids, and 40\% of irregular disks or system, versus 24\%, 21\%, and 55\% for the reference galaxies, respectively (Table~\ref{mt}).

\begin{figure*}
  	\includegraphics[width=\textwidth]{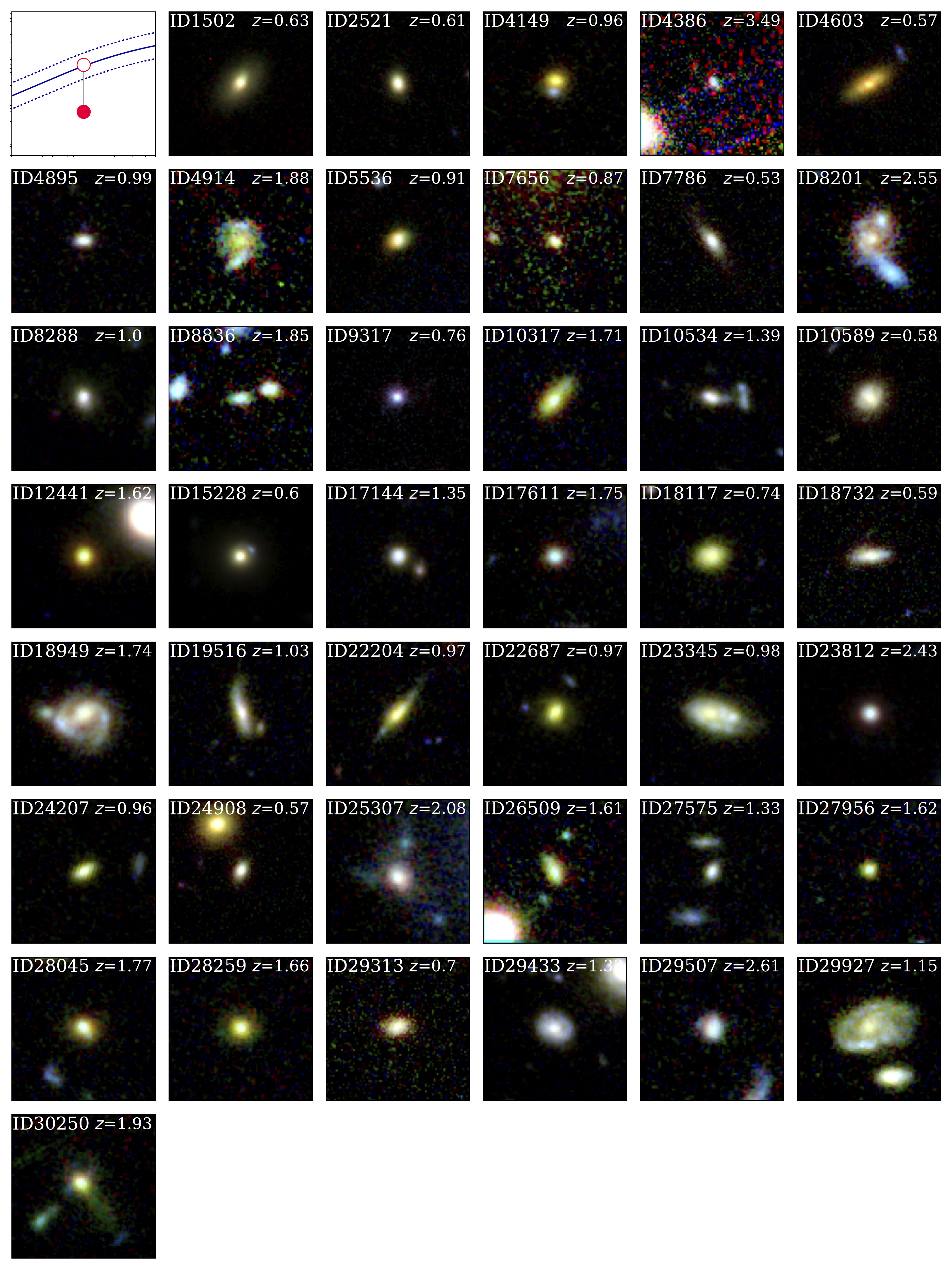}
  	\caption{ \label{image_mvm_2} Candidates that were on the MS before the downfall of star formation and that are now below the MS.}
\end{figure*}

The third group is composed of galaxies that were on the MS before and are now below it (Fig.~\ref{image_mvm_2}).
These galaxies do not span a specific stellar mass range (Fig.~\ref{rsb}).
It is composed of a large fraction of galaxies classified as spheroids (36\%) as well as pure disks (39\%).
The median stellar mass of the galaxies classified as spheroids is 2.3$\times$10$^{10}$\,M$_{\odot}$ whereas the median mass of disks is 6.8$\times$10$^{9}$\,M$_{\odot}$.
This seems to indicate that the massive galaxies experienced a morphological transformation within the MS, prior to their rapid downfall.
This morphological transformation may have been associated with the slow downfall associated to the bending of the MS \citep{Schreiber15}.
For the low mass galaxies, the fact that they are classified as disks is an indication that they have some rotation, despite the fact that being well below the MS.
The fraction of irregular systems and disks is less important, 25\% versus 44\% and 40\% for group 1 and 2, respectively, and is almost a factor of two below the proportion of irregular galaxies in the reference sample of this group (Table~\ref{mt}).
We find galaxies with low stellar mass ($\leq$10$^{10}$\,M$_{\odot}$) that departed the MS recently and lie now below it.
	
We find no significant behavior that could be different from one group to another based on the properties derived by our broad-band SED fitting procedure.
We showed that the statistical properties of our candidates point toward a recent decrease of star-formation activity in this galaxies.
However, we need spectroscopic data to definitely validate the method and its use to quickly identify such galaxies in the act. 
Furthermore, to understand the different mechanisms that are affecting the candidate galaxies of each group we need additional data in order to look for different star formation indicators, AGN activity, and gas content.
Our method can be used to pinpoint galaxies with recent variations of SFH and try to reconstruct these variations, first step to understand the evolution of galaxies in the SFR-M$_*$ plane.
Although this study focuses on galaxies undergoing a decrease of star formation activity, it could also be used, in principle, to identify movements on the MS plane due to an increase of star formation.
We find 233 candidates with  $r_{\rm{SFR}}$$>$1 and $\Delta$BIC$\geq$10 that will be explored in a future work.	
	
\section{Conclusions}
To probe and study phenomena that occur in a very short amount of time, we need a large statistic to be able to catch them in the act. 
The goal of this work is to identify galaxies that just underwent a rapid ($<$500\,Myr) and drastic downfall of their star-formation rate (more than 80\%) from broadband SED modeling since large photometric sample can provide the statistic needed to pinpoint these objects.
In this purpose, we choose to apply our method on the ZFOURGE sample to provide both statistics and large photometric coverage of the SED from UV to NIR.
\begin{itemize}
	\item We fit the whole ZFOURGE sample using the CIGALE code with two different SFHs: one normal delayed-$\tau$ SFH and one truncated SFH, and using \cite{BruzualCharlot03} models. We show that a handful of sources are better fit using the truncated SFH, that assume a recent instantaneous break in the SFH, compared to the more commonly used delayed-$\tau$ SFH.
	\item Based on the results of \cite{Ciesla16}, we define a criterion based on the quality of the fits and the output strength of the decrease of star formation activity obtained from the SED modeling. From galaxies classified as star-forming with the UVJ diagram, we select a sample of 102 candidates. 
	\item We study in details three candidates and show that the difference of fit quality between the two SFH is not due to the priors used for the SED modeling and that the normal delayed-$\tau$ struggles to model consistently with the UV rest-frame and the optical-NIR emission.
	\item Placed on the main sequence diagram, these sources lie between the lower part of the MS and the passive galaxies region. In the UVJ diagram, they lie in a region in between the passive and active regions. This confirms the fact that these sources must be in transition.
	\item To understand their differences, we define for each candidate a reference sample composed of galaxies lying in the same region in the UVJ diagram and compare their physical properties. 
	\item We compute the average properties of the reference samples associated to each candidate source.  We show that despite their common location on the UVJ diagram, the candidate galaxies are displaced in the MS diagram compared to their reference sample. They have, on average, a lower SFR and an offset in stellar mass that seems to depend on the stellar mass of the candidate.
	\item There is no specific morphology attributed to the candidates. All types are found. Their effective radius is consistent with what is expected for galaxies of the same stellar masses.
	\item Among the 102 candidates, 4 of them are detected in X-ray with a luminosity larger than 10$^{42}$\,erg/s. Another source is classified as AGN from an UV-to-NIR SED fitting decomposition. This low number imply that AGN activity is not the main responsible for the fast decrease of the star-formation activity of the candidates.
	\item Using the output of the SED fitting, we reconstruct the recent variation of SFR and show that the candidate galaxies are separated in three groups. The first one is composed of galaxies that were in the starburst regions and are now on the MS, which are thus post-starburst candidates. The second group is composed of galaxies that went from above to below the MS. The last group is composed of galaxies that were on the MS and are now below. 
\end{itemize}
In this study, we provided and used a method to blindly identify galaxies undergoing a fast diminution of their star-formation activity from their observed broadband SED.
We showed that these galaxies have indeed physical properties different from the main population and characteristic of sources in transition.
However, additional work is needed to confirm the variations in the recent SFH of these galaxies, especially using spectroscopy to look for strong Balmer absorption lines.
Ancillary data are also needed to investigate the mechanisms at play and understand the possible role of mergers, AGN, and gas content.

\begin{acknowledgements}
L.\,C. thanks M.~Boquien for useful discussions and B.~Ladjelate for his help with Python.
L.\,C. warmly thanks M.~Boquien, Y.~Roehlly and D.~Burgarella for developing the new version of CIGALE on which the present work relies on.
\end{acknowledgements}

\bibliographystyle{aa}
\bibliography{pcigale_quench}

\begin{thebibliography}{86}
\expandafter\ifx\csname natexlab\endcsname\relax\def\natexlab#1{#1}\fi

\bibitem[{{Abramson} {et~al.}(2016){Abramson}, {Gladders}, {Dressler},
  {Oemler}, {Poggianti}, \& {Vulcani}}]{Abramson16}
{Abramson}, L.~E., {Gladders}, M.~D., {Dressler}, A., {et~al.} 2016, \apj, 832,
  7

\bibitem[{{Arnouts} {et~al.}(2013){Arnouts}, {Le Floc'h}, {Chevallard},
  {Johnson}, {Ilbert}, {Treyer}, {Aussel}, {Capak}, {Sanders}, {Scoville},
  {McCracken}, {Milliard}, {Pozzetti}, \& {Salvato}}]{Arnouts13}
{Arnouts}, S., {Le Floc'h}, E., {Chevallard}, J., {et~al.} 2013, \aap, 558, A67

\bibitem[{{Ashby} {et~al.}(2013){Ashby}, {Willner}, {Fazio}, {Huang}, {Arendt},
  {Barmby}, {Barro}, {Bell}, {Bouwens}, {Cattaneo}, {Croton}, {Dav{\'e}},
  {Dunlop}, {Egami}, {Faber}, {Finlator}, {Grogin}, {Guhathakurta},
  {Hernquist}, {Hora}, {Illingworth}, {Kashlinsky}, {Koekemoer}, {Koo},
  {Labb{\'e}}, {Li}, {Lin}, {Moseley}, {Nandra}, {Newman}, {Noeske}, {Ouchi},
  {Peth}, {Rigopoulou}, {Robertson}, {Sarajedini}, {Simard}, {Smith}, {Wang},
  {Wechsler}, {Weiner}, {Wilson}, {Wuyts}, {Yamada}, \& {Yan}}]{Ashby13}
{Ashby}, M.~L.~N., {Willner}, S.~P., {Fazio}, G.~G., {et~al.} 2013, \apj, 769,
  80

\bibitem[{{Baldry} {et~al.}(2004){Baldry}, {Glazebrook}, {Brinkmann},
  {Ivezi{\'c}}, {Lupton}, {Nichol}, \& {Szalay}}]{Baldry04}
{Baldry}, I.~K., {Glazebrook}, K., {Brinkmann}, J., {et~al.} 2004, \apj, 600,
  681

\bibitem[{{Behroozi} {et~al.}(2013){Behroozi}, {Wechsler}, \&
  {Conroy}}]{Behroozi13}
{Behroozi}, P.~S., {Wechsler}, R.~H., \& {Conroy}, C. 2013, \apj, 770, 57

\bibitem[{{Boquien} {et~al.}(2012){Boquien}, {Buat}, {Boselli}, {Baes},
  {Bendo}, {Ciesla}, {Cooray}, {Cortese}, {Eales}, {Gavazzi}, {Gomez},
  {Lebouteiller}, {Pappalardo}, {Pohlen}, {Smith}, \& {Spinoglio}}]{Boquien12}
{Boquien}, M., {Buat}, V., {Boselli}, A., {et~al.} 2012, \aap, 539, A145

\bibitem[{{Boquien} {et~al.}(2014){Boquien}, {Buat}, \& {Perret}}]{Boquien14}
{Boquien}, M., {Buat}, V., \& {Perret}, V. 2014, \aap, 571, A72

\bibitem[{{Boselli} {et~al.}(2016){Boselli}, {Roehlly}, {Fossati}, {Buat},
  {Boissier}, {Boquien}, {Burgarella}, {Ciesla}, {Gavazzi}, \&
  {Serra}}]{Boselli16b}
{Boselli}, A., {Roehlly}, Y., {Fossati}, M., {et~al.} 2016, \aap, 596, A11

\bibitem[{{Brammer} {et~al.}(2012){Brammer}, {van Dokkum}, {Franx},
  {Fumagalli}, {Patel}, {Rix}, {Skelton}, {Kriek}, {Nelson}, {Schmidt},
  {Bezanson}, {da Cunha}, {Erb}, {Fan}, {F{\"o}rster Schreiber}, {Illingworth},
  {Labb{\'e}}, {Leja}, {Lundgren}, {Magee}, {Marchesini}, {McCarthy},
  {Momcheva}, {Muzzin}, {Quadri}, {Steidel}, {Tal}, {Wake}, {Whitaker}, \&
  {Williams}}]{Brammer12}
{Brammer}, G.~B., {van Dokkum}, P.~G., {Franx}, M., {et~al.} 2012, \apjs, 200,
  13

\bibitem[{{Bruzual} \& {Charlot}(2003)}]{BruzualCharlot03}
{Bruzual}, G. \& {Charlot}, S. 2003, \mnras, 344, 1000

\bibitem[{{Buat} {et~al.}(2014){Buat}, {Heinis}, {Boquien}, {Burgarella},
  {Charmandaris}, {Boissier}, {Boselli}, {Le Borgne}, \& {Morrison}}]{Buat14}
{Buat}, V., {Heinis}, S., {Boquien}, M., {et~al.} 2014, \aap, 561, A39

\bibitem[{{Calzetti} {et~al.}(2000){Calzetti}, {Armus}, {Bohlin}, {Kinney},
  {Koornneef}, \& {Storchi-Bergmann}}]{Calzetti00}
{Calzetti}, D., {Armus}, L., {Bohlin}, R.~C., {et~al.} 2000, \apj, 533, 682

\bibitem[{{Cappelluti} {et~al.}(2016){Cappelluti}, {Comastri}, {Fontana},
  {Zamorani}, {Amorin}, {Castellano}, {Merlin}, {Santini}, {Elbaz},
  {Schreiber}, {Shu}, {Wang}, {Dunlop}, {Bourne}, {Bruce}, {Buitrago},
  {Micha{\l}owski}, {Derriere}, {Ferguson}, {Faber}, \& {Vito}}]{Cappelluti16}
{Cappelluti}, N., {Comastri}, A., {Fontana}, A., {et~al.} 2016, \apj, 823, 95

\bibitem[{{Cardamone} {et~al.}(2010){Cardamone}, {van Dokkum}, {Urry},
  {Taniguchi}, {Gawiser}, {Brammer}, {Taylor}, {Damen}, {Treister}, {Cobb},
  {Bond}, {Schawinski}, {Lira}, {Murayama}, {Saito}, \&
  {Sumikawa}}]{Cardamone10}
{Cardamone}, C.~N., {van Dokkum}, P.~G., {Urry}, C.~M., {et~al.} 2010, \apjs,
  189, 270

\bibitem[{{Casey}(2012)}]{Casey12}
{Casey}, C.~M. 2012, \mnras, 425, 3094

\bibitem[{{Ciesla} {et~al.}(2016){Ciesla}, {Boselli}, {Elbaz}, {Boissier},
  {Buat}, {Charmandaris}, {Schreiber}, {B{\'e}thermin}, {Baes}, {Boquien}, {De
  Looze}, {Fern{\'a}ndez-Ontiveros}, {Pappalardo}, {Spinoglio}, \&
  {Viaene}}]{Ciesla16}
{Ciesla}, L., {Boselli}, A., {Elbaz}, D., {et~al.} 2016, \aap, 585, A43

\bibitem[{{Ciesla} {et~al.}(2015){Ciesla}, {Charmandaris}, {Georgakakis},
  {Bernhard}, {Mitchell}, {Buat}, {Elbaz}, {LeFloc'h}, {Lacey}, {Magdis}, \&
  {Xilouris}}]{Ciesla15}
{Ciesla}, L., {Charmandaris}, V., {Georgakakis}, A., {et~al.} 2015, \aap, 576,
  A10

\bibitem[{{Ciesla} {et~al.}(2017){Ciesla}, {Elbaz}, \& {Fensch}}]{Ciesla17}
{Ciesla}, L., {Elbaz}, D., \& {Fensch}, J. 2017, \aap, 608, A41

\bibitem[{{da Cunha} {et~al.}(2015){da Cunha}, {Walter}, {Smail}, {Swinbank},
  {Simpson}, {Decarli}, {Hodge}, {Weiss}, {van der Werf}, {Bertoldi},
  {Chapman}, {Cox}, {Danielson}, {Dannerbauer}, {Greve}, {Ivison}, {Karim}, \&
  {Thomson}}]{daCunha15}
{da Cunha}, E., {Walter}, F., {Smail}, I.~R., {et~al.} 2015, \apj, 806, 110

\bibitem[{{Daddi} {et~al.}(2007){Daddi}, {Dickinson}, {Morrison}, {Chary},
  {Cimatti}, {Elbaz}, {Frayer}, {Renzini}, {Pope}, {Alexander}, {Bauer},
  {Giavalisco}, {Huynh}, {Kurk}, \& {Mignoli}}]{Daddi07}
{Daddi}, E., {Dickinson}, M., {Morrison}, G., {et~al.} 2007, \apj, 670, 156

\bibitem[{{Dale} {et~al.}(2014){Dale}, {Helou}, {Magdis}, {Armus},
  {D{\'{\i}}az-Santos}, \& {Shi}}]{Dale14}
{Dale}, D.~A., {Helou}, G., {Magdis}, G.~E., {et~al.} 2014, \apj, 784, 83

\bibitem[{{Dekel} \& {Burkert}(2014)}]{DekelBurkert14}
{Dekel}, A. \& {Burkert}, A. 2014, \mnras, 438, 1870

\bibitem[{{Di Matteo} {et~al.}(2005){Di Matteo}, {Springel}, \&
  {Hernquist}}]{DiMatteo05}
{Di Matteo}, T., {Springel}, V., \& {Hernquist}, L. 2005, \nat, 433, 604

\bibitem[{{Dickinson} {et~al.}(2003){Dickinson}, {Giavalisco}, \& {GOODS
  Team}}]{Dickinson03}
{Dickinson}, M., {Giavalisco}, M., \& {GOODS Team}. 2003, in The Mass of
  Galaxies at Low and High Redshift, ed. R.~{Bender} \& A.~{Renzini}, 324

\bibitem[{{Draine} \& {Li}(2007)}]{DraineLi07}
{Draine}, B.~T. \& {Li}, A. 2007, \apj, 657, 810

\bibitem[{{Elbaz} {et~al.}(2007){Elbaz}, {Daddi}, {Le Borgne}, {Dickinson},
  {Alexander}, {Chary}, {Starck}, {Brandt}, {Kitzbichler}, {MacDonald},
  {Nonino}, {Popesso}, {Stern}, \& {Vanzella}}]{Elbaz07}
{Elbaz}, D., {Daddi}, E., {Le Borgne}, D., {et~al.} 2007, \aap, 468, 33

\bibitem[{{Elbaz} {et~al.}(2011){Elbaz}, {Dickinson}, {Hwang},
  {D{\'{\i}}az-Santos}, {Magdis}, {Magnelli}, {Le Borgne}, {Galliano},
  {Pannella}, {Chanial}, {Armus}, {Charmandaris}, {Daddi}, {Aussel}, {Popesso},
  {Kartaltepe}, {Altieri}, {Valtchanov}, {Coia}, {Dannerbauer}, {Dasyra},
  {Leiton}, {Mazzarella}, {Alexander}, {Buat}, {Burgarella}, {Chary}, {Gilli},
  {Ivison}, {Juneau}, {Le Floc'h}, {Lutz}, {Morrison}, {Mullaney}, {Murphy},
  {Pope}, {Scott}, {Brodwin}, {Calzetti}, {Cesarsky}, {Charlot}, {Dole},
  {Eisenhardt}, {Ferguson}, {F{\"o}rster Schreiber}, {Frayer}, {Giavalisco},
  {Huynh}, {Koekemoer}, {Papovich}, {Reddy}, {Surace}, {Teplitz}, {Yun}, \&
  {Wilson}}]{Elbaz11}
{Elbaz}, D., {Dickinson}, M., {Hwang}, H.~S., {et~al.} 2011, \aap, 533, A119

\bibitem[{{Elbaz} {et~al.}(2017){Elbaz}, {Leiton}, {Nagar}, {Okumura},
  {Franco}, {Schreiber}, {Pannella}, {Wang}, {Dickinson}, {Diaz-Santos},
  {Ciesla}, {Daddi}, {Bournaud}, {Magdis}, {Zhou}, \& {Rujopakarn}}]{Elbaz17}
{Elbaz}, D., {Leiton}, R., {Nagar}, N., {et~al.} 2017, ArXiv e-prints

\bibitem[{{Erben} {et~al.}(2005){Erben}, {Schirmer}, {Dietrich}, {Cordes},
  {Haberzettl}, {Hetterscheidt}, {Hildebrandt}, {Schmithuesen}, {Schneider},
  {Simon}, {Deul}, {Hook}, {Kaiser}, {Radovich}, {Benoist}, {Nonino}, {Olsen},
  {Prandoni}, {Wichmann}, {Zaggia}, {Bomans}, {Dettmar}, \&
  {Miralles}}]{Erben05}
{Erben}, T., {Schirmer}, M., {Dietrich}, J.~P., {et~al.} 2005, Astronomische
  Nachrichten, 326, 432

\bibitem[{{Fritz} {et~al.}(2006){Fritz}, {Franceschini}, \&
  {Hatziminaoglou}}]{Fritz06}
{Fritz}, J., {Franceschini}, A., \& {Hatziminaoglou}, E. 2006, \mnras, 366, 767

\bibitem[{{Gabor} \& {Bournaud}(2014)}]{Gabor14}
{Gabor}, J.~M. \& {Bournaud}, F. 2014, \mnras, 441, 1615

\bibitem[{{Gavazzi} {et~al.}(2015){Gavazzi}, {Consolandi}, {Dotti}, {Fanali},
  {Fossati}, {Fumagalli}, {Viscardi}, {Savorgnan}, {Boselli}, {Guti{\'e}rrez},
  {Hern{\'a}ndez Toledo}, {Giovanelli}, \& {Haynes}}]{Gavazzi15}
{Gavazzi}, G., {Consolandi}, G., {Dotti}, M., {et~al.} 2015, \aap, 580, A116

\bibitem[{{Giavalisco} {et~al.}(2004){Giavalisco}, {Ferguson}, {Koekemoer},
  {Dickinson}, {Alexander}, {Bauer}, {Bergeron}, {Biagetti}, {Brandt},
  {Casertano}, {Cesarsky}, {Chatzichristou}, {Conselice}, {Cristiani}, {Da
  Costa}, {Dahlen}, {de Mello}, {Eisenhardt}, {Erben}, {Fall}, {Fassnacht},
  {Fosbury}, {Fruchter}, {Gardner}, {Grogin}, {Hook}, {Hornschemeier}, {Idzi},
  {Jogee}, {Kretchmer}, {Laidler}, {Lee}, {Livio}, {Lucas}, {Madau},
  {Mobasher}, {Moustakas}, {Nonino}, {Padovani}, {Papovich}, {Park},
  {Ravindranath}, {Renzini}, {Richardson}, {Riess}, {Rosati}, {Schirmer},
  {Schreier}, {Somerville}, {Spinrad}, {Stern}, {Stiavelli}, {Strolger},
  {Urry}, {Vandame}, {Williams}, \& {Wolf}}]{Giavalisco04}
{Giavalisco}, M., {Ferguson}, H.~C., {Koekemoer}, A.~M., {et~al.} 2004, \apjl,
  600, L93

\bibitem[{{Giovannoli} {et~al.}(2011){Giovannoli}, {Buat}, {Noll},
  {Burgarella}, \& {Magnelli}}]{Giovannoli11}
{Giovannoli}, E., {Buat}, V., {Noll}, S., {Burgarella}, D., \& {Magnelli}, B.
  2011, \aap, 525, A150

\bibitem[{{Gladders} {et~al.}(2013){Gladders}, {Oemler}, {Dressler},
  {Poggianti}, {Vulcani}, \& {Abramson}}]{Gladders13}
{Gladders}, M.~D., {Oemler}, A., {Dressler}, A., {et~al.} 2013, \apj, 770, 64

\bibitem[{{Grogin} {et~al.}(2011){Grogin}, {Kocevski}, {Faber}, {Ferguson},
  {Koekemoer}, {Riess}, {Acquaviva}, {Alexander}, {Almaini}, {Ashby}, {Barden},
  {Bell}, {Bournaud}, {Brown}, {Caputi}, {Casertano}, {Cassata}, {Castellano},
  {Challis}, {Chary}, {Cheung}, {Cirasuolo}, {Conselice}, {Roshan Cooray},
  {Croton}, {Daddi}, {Dahlen}, {Dav{\'e}}, {de Mello}, {Dekel}, {Dickinson},
  {Dolch}, {Donley}, {Dunlop}, {Dutton}, {Elbaz}, {Fazio}, {Filippenko},
  {Finkelstein}, {Fontana}, {Gardner}, {Garnavich}, {Gawiser}, {Giavalisco},
  {Grazian}, {Guo}, {Hathi}, {H{\"a}ussler}, {Hopkins}, {Huang}, {Huang},
  {Jha}, {Kartaltepe}, {Kirshner}, {Koo}, {Lai}, {Lee}, {Li}, {Lotz}, {Lucas},
  {Madau}, {McCarthy}, {McGrath}, {McIntosh}, {McLure}, {Mobasher},
  {Moustakas}, {Mozena}, {Nandra}, {Newman}, {Niemi}, {Noeske}, {Papovich},
  {Pentericci}, {Pope}, {Primack}, {Rajan}, {Ravindranath}, {Reddy}, {Renzini},
  {Rix}, {Robaina}, {Rodney}, {Rosario}, {Rosati}, {Salimbeni}, {Scarlata},
  {Siana}, {Simard}, {Smidt}, {Somerville}, {Spinrad}, {Straughn}, {Strolger},
  {Telford}, {Teplitz}, {Trump}, {van der Wel}, {Villforth}, {Wechsler},
  {Weiner}, {Wiklind}, {Wild}, {Wilson}, {Wuyts}, {Yan}, \& {Yun}}]{Grogin11}
{Grogin}, N.~A., {Kocevski}, D.~D., {Faber}, S.~M., {et~al.} 2011, \apjs, 197,
  35

\bibitem[{{Guo} {et~al.}(2013){Guo}, {Zheng}, \& {Fu}}]{Guo13}
{Guo}, K., {Zheng}, X.~Z., \& {Fu}, H. 2013, \apj, 778, 23

\bibitem[{{Heinis} {et~al.}(2014){Heinis}, {Buat}, {B{\'e}thermin}, {Bock},
  {Burgarella}, {Conley}, {Cooray}, {Farrah}, {Ilbert}, {Magdis}, {Marsden},
  {Oliver}, {Rigopoulou}, {Roehlly}, {Schulz}, {Symeonidis}, {Viero}, {Xu}, \&
  {Zemcov}}]{Heinis14}
{Heinis}, S., {Buat}, V., {B{\'e}thermin}, M., {et~al.} 2014, \mnras, 437, 1268

\bibitem[{{HerMES Collaboration} {et~al.}(2012){HerMES Collaboration},
  {Oliver}, {Bock}, {Altieri}, {Amblard}, {Arumugam}, {Aussel}, {Babbedge},
  {Beelen}, {B{\'e}thermin}, {Blain}, {Boselli}, {Bridge}, {Brisbin}, {Buat},
  {Burgarella}, {Castro-Rodr{\'{\i}}guez}, {Cava}, {Chanial}, {Cirasuolo},
  {Clements}, {Conley}, {Conversi}, {Cooray}, {Dowell}, {Dubois}, {Dwek},
  {Dye}, {Eales}, {Elbaz}, {Farrah}, {Feltre}, {Ferrero}, {Fiolet}, {Fox},
  {Franceschini}, {Gear}, {Giovannoli}, {Glenn}, {Gong}, {Gonz{\'a}lez
  Solares}, {Griffin}, {Halpern}, {Harwit}, {Hatziminaoglou}, {Heinis},
  {Hurley}, {Hwang}, {Hyde}, {Ibar}, {Ilbert}, {Isaak}, {Ivison}, {Lagache},
  {Le Floc'h}, {Levenson}, {Lo Faro}, {Lu}, {Madden}, {Maffei}, {Magdis},
  {Mainetti}, {Marchetti}, {Marsden}, {Marshall}, {Mortier}, {Nguyen},
  {O'Halloran}, {Omont}, {Page}, {Panuzzo}, {Papageorgiou}, {Patel}, {Pearson},
  {P{\'e}rez-Fournon}, {Pohlen}, {Rawlings}, {Raymond}, {Rigopoulou},
  {Riguccini}, {Rizzo}, {Rodighiero}, {Roseboom}, {Rowan-Robinson},
  {S{\'a}nchez Portal}, {Schulz}, {Scott}, {Seymour}, {Shupe}, {Smith},
  {Stevens}, {Symeonidis}, {Trichas}, {Tugwell}, {Vaccari}, {Valtchanov},
  {Vieira}, {Viero}, {Vigroux}, {Wang}, {Ward}, {Wardlow}, {Wright}, {Xu}, \&
  {Zemcov}}]{Oliver12}
{HerMES Collaboration}, {Oliver}, S.~J., {Bock}, J., {et~al.} 2012, ArXiv
  e-prints

\bibitem[{{Hildebrandt} {et~al.}(2006){Hildebrandt}, {Erben}, {Dietrich},
  {Cordes}, {Haberzettl}, {Hetterscheidt}, {Schirmer}, {Schmithuesen},
  {Schneider}, {Simon}, \& {Trachternach}}]{Hildebrandt06}
{Hildebrandt}, H., {Erben}, T., {Dietrich}, J.~P., {et~al.} 2006, \aap, 452,
  1121

\bibitem[{{Hopkins} {et~al.}(2006){Hopkins}, {Hernquist}, {Cox}, {Di Matteo},
  {Robertson}, \& {Springel}}]{Hopkins06}
{Hopkins}, P.~F., {Hernquist}, L., {Cox}, T.~J., {et~al.} 2006, \apjs, 163, 1

\bibitem[{{Hsieh} {et~al.}(2012){Hsieh}, {Wang}, {Hsieh}, {Lin}, {Yan}, {Lim},
  \& {Ho}}]{Hsieh12}
{Hsieh}, B.-C., {Wang}, W.-H., {Hsieh}, C.-C., {et~al.} 2012, \apjs, 203, 23

\bibitem[{{Huertas-Company} {et~al.}(2016){Huertas-Company}, {Bernardi},
  {P{\'e}rez-Gonz{\'a}lez}, {Ashby}, {Barro}, {Conselice}, {Daddi}, {Dekel},
  {Dimauro}, {Faber}, {Grogin}, {Kartaltepe}, {Kocevski}, {Koekemoer}, {Koo},
  {Mei}, \& {Shankar}}]{HuertasCompany16}
{Huertas-Company}, M., {Bernardi}, M., {P{\'e}rez-Gonz{\'a}lez}, P.~G.,
  {et~al.} 2016, ArXiv e-prints

\bibitem[{{Ilbert} {et~al.}(2015){Ilbert}, {Arnouts}, {Le Floc'h}, {Aussel},
  {Bethermin}, {Capak}, {Hsieh}, {Kajisawa}, {Karim}, {Le F{\`e}vre}, {Lee},
  {Lilly}, {McCracken}, {Michel-Dansac}, {Moutard}, {Renzini}, {Salvato},
  {Sanders}, {Scoville}, {Sheth}, {Silverman}, {Smol{\v c}i{\'c}}, {Taniguchi},
  \& {Tresse}}]{Ilbert15}
{Ilbert}, O., {Arnouts}, S., {Le Floc'h}, E., {et~al.} 2015, \aap, 579, A2

\bibitem[{{Koekemoer} {et~al.}(2011){Koekemoer}, {Faber}, {Ferguson}, {Grogin},
  {Kocevski}, {Koo}, {Lai}, {Lotz}, {Lucas}, {McGrath}, {Ogaz}, {Rajan},
  {Riess}, {Rodney}, {Strolger}, {Casertano}, {Castellano}, {Dahlen},
  {Dickinson}, {Dolch}, {Fontana}, {Giavalisco}, {Grazian}, {Guo}, {Hathi},
  {Huang}, {van der Wel}, {Yan}, {Acquaviva}, {Alexander}, {Almaini}, {Ashby},
  {Barden}, {Bell}, {Bournaud}, {Brown}, {Caputi}, {Cassata}, {Challis},
  {Chary}, {Cheung}, {Cirasuolo}, {Conselice}, {Roshan Cooray}, {Croton},
  {Daddi}, {Dav{\'e}}, {de Mello}, {de Ravel}, {Dekel}, {Donley}, {Dunlop},
  {Dutton}, {Elbaz}, {Fazio}, {Filippenko}, {Finkelstein}, {Frazer}, {Gardner},
  {Garnavich}, {Gawiser}, {Gruetzbauch}, {Hartley}, {H{\"a}ussler},
  {Herrington}, {Hopkins}, {Huang}, {Jha}, {Johnson}, {Kartaltepe},
  {Khostovan}, {Kirshner}, {Lani}, {Lee}, {Li}, {Madau}, {McCarthy},
  {McIntosh}, {McLure}, {McPartland}, {Mobasher}, {Moreira}, {Mortlock},
  {Moustakas}, {Mozena}, {Nandra}, {Newman}, {Nielsen}, {Niemi}, {Noeske},
  {Papovich}, {Pentericci}, {Pope}, {Primack}, {Ravindranath}, {Reddy},
  {Renzini}, {Rix}, {Robaina}, {Rosario}, {Rosati}, {Salimbeni}, {Scarlata},
  {Siana}, {Simard}, {Smidt}, {Snyder}, {Somerville}, {Spinrad}, {Straughn},
  {Telford}, {Teplitz}, {Trump}, {Vargas}, {Villforth}, {Wagner}, {Wandro},
  {Wechsler}, {Weiner}, {Wiklind}, {Wild}, {Wilson}, {Wuyts}, \&
  {Yun}}]{Koekemoer11}
{Koekemoer}, A.~M., {Faber}, S.~M., {Ferguson}, H.~C., {et~al.} 2011, \apjs,
  197, 36

\bibitem[{{Komatsu} {et~al.}(2011){Komatsu}, {Smith}, {Dunkley}, {Bennett},
  {Gold}, {Hinshaw}, {Jarosik}, {Larson}, {Nolta}, {Page}, {Spergel},
  {Halpern}, {Hill}, {Kogut}, {Limon}, {Meyer}, {Odegard}, {Tucker}, {Weiland},
  {Wollack}, \& {Wright}}]{Komatsu11}
{Komatsu}, E., {Smith}, K.~M., {Dunkley}, J., {et~al.} 2011, \apjs, 192, 18

\bibitem[{{Kriek} {et~al.}(2009){Kriek}, {van Dokkum}, {Labb{\'e}}, {Franx},
  {Illingworth}, {Marchesini}, \& {Quadri}}]{Kriek09}
{Kriek}, M., {van Dokkum}, P.~G., {Labb{\'e}}, I., {et~al.} 2009, \apj, 700,
  221

\bibitem[{{Lee} {et~al.}(2010){Lee}, {Ferguson}, {Somerville}, {Wiklind}, \&
  {Giavalisco}}]{Lee10}
{Lee}, S.-K., {Ferguson}, H.~C., {Somerville}, R.~S., {Wiklind}, T., \&
  {Giavalisco}, M. 2010, \apj, 725, 1644

\bibitem[{{Liddle}(2004)}]{Liddle04}
{Liddle}, A.~R. 2004, \mnras, 351, L49

\bibitem[{{Lutz} {et~al.}(2011){Lutz}, {Poglitsch}, {Altieri}, {Andreani},
  {Aussel}, {Berta}, {Bongiovanni}, {Brisbin}, {Cava}, {Cepa}, {Cimatti},
  {Daddi}, {Dominguez-Sanchez}, {Elbaz}, {F{\"o}rster Schreiber}, {Genzel},
  {Grazian}, {Gruppioni}, {Harwit}, {Le Floc'h}, {Magdis}, {Magnelli},
  {Maiolino}, {Nordon}, {P{\'e}rez Garc{\'{\i}}a}, {Popesso}, {Pozzi},
  {Riguccini}, {Rodighiero}, {Saintonge}, {Sanchez Portal}, {Santini}, {Shao},
  {Sturm}, {Tacconi}, {Valtchanov}, {Wetzstein}, \& {Wieprecht}}]{Lutz11}
{Lutz}, D., {Poglitsch}, A., {Altieri}, B., {et~al.} 2011, \aap, 532, A90

\bibitem[{{Ma} {et~al.}(2015){Ma}, {Gonzalez}, {Spilker}, {Strandet}, {Ashby},
  {Aravena}, {B{\'e}thermin}, {Bothwell}, {de Breuck}, {Brodwin}, {Chapman},
  {Fassnacht}, {Greve}, {Gullberg}, {Hezaveh}, {Malkan}, {Marrone},
  {Saliwanchik}, {Vieira}, {Weiss}, \& {Welikala}}]{Ma15}
{Ma}, J., {Gonzalez}, A.~H., {Spilker}, J.~S., {et~al.} 2015, \apj, 812, 88

\bibitem[{{Magdis} {et~al.}(2012){Magdis}, {Daddi}, {B{\'e}thermin}, {Sargent},
  {Elbaz}, {Pannella}, {Dickinson}, {Dannerbauer}, {da Cunha}, {Walter},
  {Rigopoulou}, {Charmandaris}, {Hwang}, \& {Kartaltepe}}]{Magdis12}
{Magdis}, G.~E., {Daddi}, E., {B{\'e}thermin}, M., {et~al.} 2012, \apj, 760, 6

\bibitem[{{Mancuso} {et~al.}(2016){Mancuso}, {Lapi}, {Shi}, {Cai},
  {Gonzalez-Nuevo}, {B{\'e}thermin}, \& {Danese}}]{Mancuso16}
{Mancuso}, C., {Lapi}, A., {Shi}, J., {et~al.} 2016, \apj, 833, 152

\bibitem[{{Maraston}(2005)}]{Maraston05}
{Maraston}, C. 2005, \mnras, 362, 799

\bibitem[{{Maraston} {et~al.}(2010){Maraston}, {Pforr}, {Renzini}, {Daddi},
  {Dickinson}, {Cimatti}, \& {Tonini}}]{Maraston10}
{Maraston}, C., {Pforr}, J., {Renzini}, A., {et~al.} 2010, \mnras, 407, 830

\bibitem[{{Noeske} {et~al.}(2007){Noeske}, {Weiner}, {Faber}, {Papovich},
  {Koo}, {Somerville}, {Bundy}, {Conselice}, {Newman}, {Schiminovich}, {Le
  Floc'h}, {Coil}, {Rieke}, {Lotz}, {Primack}, {Barmby}, {Cooper}, {Davis},
  {Ellis}, {Fazio}, {Guhathakurta}, {Huang}, {Kassin}, {Martin}, {Phillips},
  {Rich}, {Small}, {Willmer}, \& {Wilson}}]{Noeske07}
{Noeske}, K.~G., {Weiner}, B.~J., {Faber}, S.~M., {et~al.} 2007, \apjl, 660,
  L43

\bibitem[{{Noll} {et~al.}(2009){Noll}, {Burgarella}, {Giovannoli}, {Buat},
  {Marcillac}, \& {Mu{\~n}oz-Mateos}}]{Noll09}
{Noll}, S., {Burgarella}, D., {Giovannoli}, E., {et~al.} 2009, \aap, 507, 1793

\bibitem[{{Nonino} {et~al.}(2009){Nonino}, {Dickinson}, {Rosati}, {Grazian},
  {Reddy}, {Cristiani}, {Giavalisco}, {Kuntschner}, {Vanzella}, {Daddi},
  {Fosbury}, \& {Cesarsky}}]{Nonino09}
{Nonino}, M., {Dickinson}, M., {Rosati}, P., {et~al.} 2009, \apjs, 183, 244

\bibitem[{{Pacifici} {et~al.}(2013){Pacifici}, {Kassin}, {Weiner}, {Charlot},
  \& {Gardner}}]{Pacifici13}
{Pacifici}, C., {Kassin}, S.~A., {Weiner}, B., {Charlot}, S., \& {Gardner},
  J.~P. 2013, \apjl, 762, L15

\bibitem[{{Pacifici} {et~al.}(2016){Pacifici}, {Kassin}, {Weiner}, {Holden},
  {Gardner}, {Faber}, {Ferguson}, {Koo}, {Primack}, {Bell}, {Dekel}, {Gawiser},
  {Giavalisco}, {Rafelski}, {Simons}, {Barro}, {Croton}, {Dav{\'e}}, {Fontana},
  {Grogin}, {Koekemoer}, {Lee}, {Salmon}, {Somerville}, \&
  {Behroozi}}]{Pacifici16}
{Pacifici}, C., {Kassin}, S.~A., {Weiner}, B.~J., {et~al.} 2016, \apj, 832, 79

\bibitem[{{Pannella} {et~al.}(2009){Pannella}, {Carilli}, {Daddi}, {McCracken},
  {Owen}, {Renzini}, {Strazzullo}, {Civano}, {Koekemoer}, {Schinnerer},
  {Scoville}, {Smol{\v c}i{\'c}}, {Taniguchi}, {Aussel}, {Kneib}, {Ilbert},
  {Mellier}, {Salvato}, {Thompson}, \& {Willott}}]{Pannella09}
{Pannella}, M., {Carilli}, C.~L., {Daddi}, E., {et~al.} 2009, \apjl, 698, L116

\bibitem[{{Papovich} {et~al.}(2001){Papovich}, {Dickinson}, \&
  {Ferguson}}]{Papovich01}
{Papovich}, C., {Dickinson}, M., \& {Ferguson}, H.~C. 2001, \apj, 559, 620

\bibitem[{{Pforr} {et~al.}(2012){Pforr}, {Maraston}, \& {Tonini}}]{Pforr12}
{Pforr}, J., {Maraston}, C., \& {Tonini}, C. 2012, \mnras, 422, 3285

\bibitem[{{Retzlaff} {et~al.}(2010){Retzlaff}, {Rosati}, {Dickinson},
  {Vandame}, {Rit{\'e}}, {Nonino}, {Cesarsky}, \& {GOODS Team}}]{Retzlaff10}
{Retzlaff}, J., {Rosati}, P., {Dickinson}, M., {et~al.} 2010, \aap, 511, A50

\bibitem[{{Rodighiero} {et~al.}(2011){Rodighiero}, {Daddi}, {Baronchelli},
  {Cimatti}, {Renzini}, {Aussel}, {Popesso}, {Lutz}, {Andreani}, {Berta},
  {Cava}, {Elbaz}, {Feltre}, {Fontana}, {F{\"o}rster Schreiber},
  {Franceschini}, {Genzel}, {Grazian}, {Gruppioni}, {Ilbert}, {Le Floch},
  {Magdis}, {Magliocchetti}, {Magnelli}, {Maiolino}, {McCracken}, {Nordon},
  {Poglitsch}, {Santini}, {Pozzi}, {Riguccini}, {Tacconi}, {Wuyts}, \&
  {Zamorani}}]{Rodighiero11}
{Rodighiero}, G., {Daddi}, E., {Baronchelli}, I., {et~al.} 2011, \apjl, 739,
  L40

\bibitem[{{Roehlly} {et~al.}(2014){Roehlly}, {Burgarella}, {Buat}, {Boquien},
  {Ciesla}, \& {Heinis}}]{Roehlly14}
{Roehlly}, Y., {Burgarella}, D., {Buat}, V., {et~al.} 2014, in Astronomical
  Society of the Pacific Conference Series, Vol. 485, Astronomical Data
  Analysis Software and Systems XXIII, ed. N.~{Manset} \& P.~{Forshay}, 347

\bibitem[{{Salmi} {et~al.}(2012){Salmi}, {Daddi}, {Elbaz}, {Sargent},
  {Dickinson}, {Renzini}, {Bethermin}, \& {Le Borgne}}]{Salmi12}
{Salmi}, F., {Daddi}, E., {Elbaz}, D., {et~al.} 2012, \apjl, 754, L14

\bibitem[{{Salpeter}(1955)}]{Salpeter55}
{Salpeter}, E.~E. 1955, \apj, 121, 161

\bibitem[{{Sargent} {et~al.}(2014){Sargent}, {Daddi}, {B{\'e}thermin},
  {Aussel}, {Magdis}, {Hwang}, {Juneau}, {Elbaz}, \& {da Cunha}}]{Sargent14}
{Sargent}, M.~T., {Daddi}, E., {B{\'e}thermin}, M., {et~al.} 2014, \apj, 793,
  19

\bibitem[{{Schreiber} {et~al.}(2015){Schreiber}, {Pannella}, {Elbaz},
  {B{\'e}thermin}, {Inami}, {Dickinson}, {Magnelli}, {Wang}, {Aussel}, {Daddi},
  {Juneau}, {Shu}, {Sargent}, {Buat}, {Faber}, {Ferguson}, {Giavalisco},
  {Koekemoer}, {Magdis}, {Morrison}, {Papovich}, {Santini}, \&
  {Scott}}]{Schreiber15}
{Schreiber}, C., {Pannella}, M., {Elbaz}, D., {et~al.} 2015, \aap, 575, A74

\bibitem[{{Schreiber} {et~al.}(2017){Schreiber}, {Pannella}, {Leiton}, {Elbaz},
  {Wang}, {Okumura}, \& {Labb{\'e}}}]{Schreiber17}
{Schreiber}, C., {Pannella}, M., {Leiton}, R., {et~al.} 2017, \aap, 599, A134

\bibitem[{{Scoville} {et~al.}(2017){Scoville}, {Lee}, {Vanden Bout},
  {Diaz-Santos}, {Sanders}, {Darvish}, {Bongiorno}, {Casey}, {Murchikova},
  {Koda}, {Capak}, {Vlahakis}, {Ilbert}, {Sheth}, {Morokuma-Matsui}, {Ivison},
  {Aussel}, {Laigle}, {McCracken}, {Armus}, {Pope}, {Toft}, \&
  {Masters}}]{Scoville16}
{Scoville}, N., {Lee}, N., {Vanden Bout}, P., {et~al.} 2017, \apj, 837, 150

\bibitem[{{Simha} {et~al.}(2014){Simha}, {Weinberg}, {Conroy}, {Dave},
  {Fardal}, {Katz}, \& {Oppenheimer}}]{Simha14}
{Simha}, V., {Weinberg}, D.~H., {Conroy}, C., {et~al.} 2014, ArXiv e-prints

\bibitem[{{Speagle} {et~al.}(2014){Speagle}, {Steinhardt}, {Capak}, \&
  {Silverman}}]{Speagle14}
{Speagle}, J.~S., {Steinhardt}, C.~L., {Capak}, P.~L., \& {Silverman}, J.~D.
  2014, \apjs, 214, 15

\bibitem[{{Spitler} {et~al.}(2014){Spitler}, {Straatman}, {Labb{\'e}},
  {Glazebrook}, {Tran}, {Kacprzak}, {Quadri}, {Papovich}, {Persson}, {van
  Dokkum}, {Allen}, {Kawinwanichakij}, {Kelson}, {McCarthy}, {Mehrtens},
  {Monson}, {Nanayakkara}, {Rees}, {Tilvi}, \& {Tomczak}}]{Spitler14}
{Spitler}, L.~R., {Straatman}, C.~M.~S., {Labb{\'e}}, I., {et~al.} 2014, \apjl,
  787, L36

\bibitem[{{Stanley} {et~al.}(2015){Stanley}, {Harrison}, {Alexander},
  {Swinbank}, {Aird}, {Del Moro}, {Hickox}, \& {Mullaney}}]{Stanley15}
{Stanley}, F., {Harrison}, C.~M., {Alexander}, D.~M., {et~al.} 2015, \mnras,
  453, 591

\bibitem[{{Straatman} {et~al.}(2014){Straatman}, {Labb{\'e}}, {Spitler},
  {Allen}, {Altieri}, {Brammer}, {Dickinson}, {van Dokkum}, {Inami},
  {Glazebrook}, {Kacprzak}, {Kawinwanichakij}, {Kelson}, {McCarthy},
  {Mehrtens}, {Monson}, {Murphy}, {Papovich}, {Persson}, {Quadri}, {Rees},
  {Tomczak}, {Tran}, \& {Tilvi}}]{Straatman14}
{Straatman}, C.~M.~S., {Labb{\'e}}, I., {Spitler}, L.~R., {et~al.} 2014, \apjl,
  783, L14

\bibitem[{{Straatman} {et~al.}(2016){Straatman}, {Spitler}, {Quadri},
  {Labb{\'e}}, {Glazebrook}, {Persson}, {Papovich}, {Tran}, {Brammer},
  {Cowley}, {Tomczak}, {Nanayakkara}, {Alcorn}, {Allen}, {Broussard}, {van
  Dokkum}, {Forrest}, {van Houdt}, {Kacprzak}, {Kawinwanichakij}, {Kelson},
  {Lee}, {McCarthy}, {Mehrtens}, {Monson}, {Murphy}, {Rees}, {Tilvi}, \&
  {Whitaker}}]{Straatman16}
{Straatman}, C.~M.~S., {Spitler}, L.~R., {Quadri}, R.~F., {et~al.} 2016, \apj,
  830, 51

\bibitem[{{Tacchella} {et~al.}(2016){Tacchella}, {Dekel}, {Carollo},
  {Ceverino}, {DeGraf}, {Lapiner}, {Mandelker}, \& {Primack
  Joel}}]{Tacchella16}
{Tacchella}, S., {Dekel}, A., {Carollo}, C.~M., {et~al.} 2016, \mnras, 457,
  2790

\bibitem[{{Tomczak} {et~al.}(2016){Tomczak}, {Quadri}, {Tran}, {Labb{\'e}},
  {Straatman}, {Papovich}, {Glazebrook}, {Allen}, {Brammer}, {Cowley},
  {Dickinson}, {Elbaz}, {Inami}, {Kacprzak}, {Morrison}, {Nanayakkara},
  {Persson}, {Rees}, {Salmon}, {Schreiber}, {Spitler}, \&
  {Whitaker}}]{Tomczak16}
{Tomczak}, A.~R., {Quadri}, R.~F., {Tran}, K.-V.~H., {et~al.} 2016, \apj, 817,
  118

\bibitem[{{Tomczak} {et~al.}(2014){Tomczak}, {Quadri}, {Tran}, {Labb{\'e}},
  {Straatman}, {Papovich}, {Glazebrook}, {Allen}, {Brammer}, {Kacprzak},
  {Kawinwanichakij}, {Kelson}, {McCarthy}, {Mehrtens}, {Monson}, {Persson},
  {Spitler}, {Tilvi}, \& {van Dokkum}}]{Tomczak14}
{Tomczak}, A.~R., {Quadri}, R.~F., {Tran}, K.-V.~H., {et~al.} 2014, \apj, 783,
  85

\bibitem[{{van der Wel} {et~al.}(2014){van der Wel}, {Franx}, {van Dokkum},
  {Skelton}, {Momcheva}, {Whitaker}, {Brammer}, {Bell}, {Rix}, {Wuyts},
  {Ferguson}, {Holden}, {Barro}, {Koekemoer}, {Chang}, {McGrath},
  {H{\"a}ussler}, {Dekel}, {Behroozi}, {Fumagalli}, {Leja}, {Lundgren},
  {Maseda}, {Nelson}, {Wake}, {Patel}, {Labb{\'e}}, {Faber}, {Grogin}, \&
  {Kocevski}}]{VanderWel14}
{van der Wel}, A., {Franx}, M., {van Dokkum}, P.~G., {et~al.} 2014, \apj, 788,
  28

\bibitem[{{Whitaker} {et~al.}(2014){Whitaker}, {Franx}, {Leja}, {van Dokkum},
  {Henry}, {Skelton}, {Fumagalli}, {Momcheva}, {Brammer}, {Labb{\'e}},
  {Nelson}, \& {Rigby}}]{Whitaker14}
{Whitaker}, K.~E., {Franx}, M., {Leja}, J., {et~al.} 2014, \apj, 795, 104

\bibitem[{{Whitaker} {et~al.}(2011){Whitaker}, {Labb{\'e}}, {van Dokkum},
  {Brammer}, {Kriek}, {Marchesini}, {Quadri}, {Franx}, {Muzzin}, {Williams},
  {Bezanson}, {Illingworth}, {Lee}, {Lundgren}, {Nelson}, {Rudnick}, {Tal}, \&
  {Wake}}]{Whitaker11}
{Whitaker}, K.~E., {Labb{\'e}}, I., {van Dokkum}, P.~G., {et~al.} 2011, \apj,
  735, 86

\bibitem[{{Wuyts} {et~al.}(2011){Wuyts}, {F{\"o}rster Schreiber}, {van der
  Wel}, {Magnelli}, {Guo}, {Genzel}, {Lutz}, {Aussel}, {Barro}, {Berta},
  {Cava}, {Graci{\'a}-Carpio}, {Hathi}, {Huang}, {Kocevski}, {Koekemoer},
  {Lee}, {Le Floc'h}, {McGrath}, {Nordon}, {Popesso}, {Pozzi}, {Riguccini},
  {Rodighiero}, {Saintonge}, \& {Tacconi}}]{Wuyts11}
{Wuyts}, S., {F{\"o}rster Schreiber}, N.~M., {van der Wel}, A., {et~al.} 2011,
  \apj, 742, 96

\bibitem[{{Wuyts} {et~al.}(2008){Wuyts}, {Labb{\'e}}, {F{\"o}rster Schreiber},
  {Franx}, {Rudnick}, {Brammer}, \& {van Dokkum}}]{Wuyts08}
{Wuyts}, S., {Labb{\'e}}, I., {F{\"o}rster Schreiber}, N.~M., {et~al.} 2008,
  \apj, 682, 985

\end{thebibliography}

\end{document}